\documentclass[prb,twocolumn,showpacs,superscriptaddress,floatfix]{revtex4-1}
\usepackage{graphicx}
\usepackage{amssymb}
\usepackage{amsmath}
\usepackage{xspace}
\usepackage{dcolumn}
\usepackage{bm}
\usepackage{color}
\usepackage[extra]{tipa}

\setcitestyle{square,numbers}

\newfont{\tensy}{cmsy10}

\newcommand{\ie}[0]{i.e.\@\xspace}
\newcommand{\eg}[0]{e.g.\@\xspace}

\newcommand{\rmi}{\text{i}}
\newcommand{\UP}[0]{\uparrow}
\newcommand{\DO}[0]{\downarrow}

\newcommand{\on}{\hat{n}}

\newcommand{\bit}{\begin{itemize}}
\newcommand{\eit}{\end{itemize}}

\newcommand{\om}[0]{\omega}

\newcommand{\kF}{k_\text{F}}
\newcommand{\kB}{k_\text{B}}
\newcommand{\nag}{{\phantom{\dag}}}

\newcommand{\Kr}{K_\rho}
\newcommand{\Ks}{K_\sigma}

\newcommand{\lc}{\lambda_{c}}

\newcommand{\las}[0]{\langle}
\newcommand{\ras}[0]{\rangle}

\renewcommand{\tilde}[1]{\widetilde{#1}}

\begin{document}


\title{Finite-size effects in Luther-Emery phases of Holstein and Hubbard models}

\author{J. Greitemann}
\affiliation{Department of Physics, Arnold Sommerfeld Center for Theoretical
  Physics and Center for NanoScience, Ludwig-Maximilians-Universit\"at M\"unchen,
  80333 Munich, Germany}

\affiliation{Institute for Theoretical Solid State Physics, JARA-FIT,\\
 and JARA-HPC, RWTH Aachen University, 52056 Aachen, Germany}

\author{S. Hesselmann}
\affiliation{Institute for Theoretical Solid State Physics, JARA-FIT,\\
 and JARA-HPC, RWTH Aachen University, 52056 Aachen, Germany}

\author{S. Wessel}
\affiliation{Institute for Theoretical Solid State Physics, JARA-FIT,\\
 and JARA-HPC, RWTH Aachen University, 52056 Aachen, Germany}

\author{F. F. Assaad}

\author{M. Hohenadler}
\affiliation{\mbox{Institut f\"ur Theoretische Physik und Astrophysik,
    Universit\"at W\"urzburg, 97074 W\"urzburg, Germany}}

\begin{abstract}
  The one-dimensional Holstein model and its generalizations have been
  studied extensively to understand the effects of electron-phonon
  interaction.  The half-filled case is of particular interest, as it
  describes a transition from a metallic phase with a spin gap due to
  attractive backscattering to a Peierls insulator with
  charge-density-wave (CDW) order. Our quantum Monte Carlo results support
  the existence of a metallic phase with dominant power-law charge
  correlations, as described by the Luther-Emery fixed point. We demonstrate
  that for Holstein and also for purely fermionic models the
  spin gap significantly complicates finite-size numerical studies, and
  explains inconsistent previous results for Luttinger parameters and phase
  boundaries.  On the other hand, no such complications arise in spinless models.
  The correct low-energy theory of the spinful Holstein model is argued to be
  that of singlet bipolarons with a repulsive, mutual interaction. This
  picture naturally explains the existence of a metallic phase, but also
  implies that gapless Luttinger liquid theory is not applicable.
\end{abstract}

\date{\today}

\pacs{71.30.+h, 02.70.Ss, 64.60.an, 71.10.Pm}

\maketitle

\section{Introduction}\label{sec:intro}

The interaction of charge carriers with the crystal lattice is a fundamental
ingredient for the description of materials. In addition to its
role for superconductivity, electron-phonon coupling manifests itself in
phenomena such as the Peierls transition \cite{Peierls} and polaron
formation \cite{Landau_Polaron}.  Electron-phonon coupling appears to play an
important role in a number of strongly correlated materials such as the
manganites \cite{David_AiP}. From a theoretical point of view, even a single
electron coupled to quantum phonons represents a complex many-body
problem \cite{0034-4885-72-6-066501}, and systems with a finite band filling
are even more demanding.

Much of our knowledge about the effects of electron-phonon coupling
in low-dimensional systems comes from investigations of simplified
microscopic models. Among these models, the Holstein model \cite{Ho59a} and its
generalizations play a central role. The focus on
Holstein-type models is a consequence of its relative simplicity, namely a
density-displacement coupling and Einstein phonons, which makes the models
amenable to exact numerical methods such as quantum Monte Carlo (QMC) or the
density-matrix renormalization group (DMRG); for a
review of existing work see Ref.~\onlinecite{PhysRevB87.075149}. In the
one-dimensional (1D) case considered here, Holstein-type models
have also been studied with the bosonization method in
combination with the renormalization group (RG) \cite{PhysRevB.71.205113,Ba.Bo.07}.

A critical look at the available literature reveals that,
despite the substantial body of results, the understanding of the half-filled
Holstein and Holstein-Hubbard models is incomplete. Apart from discrepancies
in the critical values
obtained with different
numerical methods, there is a long-standing
argument regarding the existence of a metallic phase.
Whereas more recent large-scale numerical
studies \cite{ClHa05,hardikar:245103,0295-5075-84-5-57001,1742-6596-200-1-012031}
support an extended metallic region as a result of quantum
lattice fluctuations, similar to the spinless Holstein
model \cite{BuMKHa98,Hohenadler06,Ej.Fe.09}, earlier
numerical and analytical results imply an insulating ground state for any
nonzero electron-phonon coupling \cite{Hirsch83a,PhysRevB.71.205113,Ba.Bo.07}.
A recent functional RG study confirmed the existence of an
extended metallic region \cite{Barkim2015}. Complications
also arise when trying to understand numerical results in terms of Luttinger
liquid theory. Most notably, the numerically determined values of the Luttinger
parameter $\Kr$ \cite{ClHa05,hardikar:245103,0295-5075-84-5-57001} conflict with the observed behavior of the
correlation functions \cite{PhysRevB87.075149,tam:161103,PhysRevB.84.165123,TeArAo05,PhysRevB.76.155114}.
The situation is further complicated by the fact that
the electron-phonon interaction gives rise to a spin gap
\cite{PhysRevB87.075149}, so that the low-energy physics is
described by the Luther-Emery fixed point \cite{Lu.Em.74}.

Here, we reveal the key role of the spin degrees of freedom for
the observed inconsistencies.
In addition to surveying previous results, we present new results for
Holstein and purely fermionic models. A comparison between
spinful and spinless Holstein models allows us to separate the effects of
backscattering from retardation effects related to the lattice dynamics.
On the other hand, transitions from a Luttinger
or Luther-Emery liquid to a CDW phase can also be investigated
in numerically more accessible fermionic models, for which a
quantitative comparison between theory and simulation is possible.
The central findings of this work are as follows. (i) Previous
claims for the absence of a metallic phase \cite{Hirsch83a} are shown to be
unfounded. Instead, the spinful Holstein model seems to support a Peierls
transition at a nonzero critical electron-phonon coupling. (ii) The metallic
phase of the spinful Holstein model is characterized by power-law (dominant)
charge and (subdominant) pairing correlations, and
exponentially suppressed spin correlations. (iii) The spin gap caused by
backscattering lies at the heart of the conflicts between numerics and
theory. In particular, the expected
Luther-Emery behavior is only observed on length scales larger than the
inverse spin gap. (iv) The pairing of
electrons into spin singlets (bipolarons) provides a connection to the
physics of the better understood spinless Holstein model, and suggests that
the correct low-energy picture of the Holstein model is that of hard-core bosons with
repulsive interactions. The latter problem can be solved exactly, 
and provides a natural explanation for the
existence of an extended metallic region in the Holstein model.

The paper is organized as follows. In Sec.~\ref{sec:model}, we
introduce the relevant models. The QMC methods 
are briefly explained in Sec.~\ref{sec:method}. In Sec.~\ref{sec:holstein},
we discuss the spinful and spinless Holstein models, including correlation
functions, Luttinger parameters, and the charge susceptibility. In
Sec.~\ref{sec:simhubb}, we reveal important similarities with suitable fermionic
models that capture the metal-insulator
transition. Section~\ref{sec:discussion} gives a discussion of our and
previous results and the correct low-energy theory of the Holstein model.
Finally, Section~\ref{sec:conclusions} contains our conclusions, and the
Appendix gives the derivation of the scaling behavior of the charge susceptibility.

\section{Models}\label{sec:model}

The 1D Holstein model is defined as \cite{Ho59a}
\begin{align}\label{eq:model-holsteinspinful}
  \hat{H}  
  &=
  -t\sum_{i\sigma} \left( c^{\dag}_{i\sigma} c^\nag_{i+1\sigma} + \text{H.c.} \right)
  \\\nonumber
  &\quad+ \sum_{i} \left(\mbox{$\frac{1}{2M}$} \hat{P}_{i}^2 + \mbox{$\frac{K}{2}$}
    \hat{Q}_{i}^2 \right)
  - g \sum_{i} \hat{Q}_{i} \left( \hat{n}_{i}-1\right) 
  \,.
\end{align}
The first term describes the hopping of electrons between neighboring lattice
sites with amplitude $t$. The second term describes the lattice
degrees of freedom in the harmonic approximation; the phonon frequency
is given by $\om_0=\sqrt{K/M}$, $\hat{Q}_i$ ($\hat{P}_i$) is the
lattice displacement (momentum) at site $i$. The electron-phonon interaction
described by the third term couples the electron density at site
$i$ with the lattice distortion at the same  site. The electron density operator is defined as $\hat{n}_{i\sigma} =
c^{\dag}_{i\sigma } c^\nag_{i\sigma }$, and we have $\on_i =
\sum_\sigma\on_{i\sigma}$.  The dimensionless parameter $\lambda=g^2/(4Kt)$ is
a useful measure for the electron-phonon coupling strength. 
A common alternative notation for the interaction term is 
 $-\overline{g} \sum_{i} (b^{\dag}_i+b_i) (
\hat{n}_{i}-1)$ where $g=\overline{g}\sqrt{2M\om_0}$ and $\lambda=
\overline{g}^2/(2\om_0)$.
Equation~(\ref{eq:model-holsteinspinful}) has been generalized to
the Holstein-Hubbard model \cite{vdL.Be.Va.95} with an additional repulsive
interaction $U \sum_i \on_{i\UP} \on_{i\DO}$, see
Ref.~\onlinecite{PhysRevB87.075149} for a review.

The spinless Holstein model \cite{Hirsch83a}
\begin{align}\label{eq:model-holsteinspinless}
  \hat{H}  
  =
  &-t\sum_{i} \left( c^{\dag}_{i} c^\nag_{i+1} + \text{H.c.} \right)
  \\\nonumber
  &+ \sum_{i} \left(\mbox{$\frac{1}{2M}$} \hat{P}_{i}^2 + \mbox{$\frac{K}{2}$}
    \hat{Q}_{i}^2 \right)
  - g \sum_{i} \hat{Q}_{i} \left( \hat{n}_{i}-0.5\right)\,, 
\end{align}
with $\on_i = c^\dag_i c^\nag_i$, captures much of the physics of
Eq.~(\ref{eq:model-holsteinspinful}), and will provide important insights.

We further consider the $U$-$V$ extended Hubbard model
\begin{align}\label{eq:model-hubbard}
  \hat{H}  
  =
  &-t\sum_{i\sigma} \left( c^{\dag}_{i\sigma} c^\nag_{i+1\sigma} + \text{H.c.}
  \right)
  \\\nonumber
  &+ U \sum_i \on_{i\UP} \on_{i\DO}
  + V \sum_i \on_{i} \on_{i+1}
\end{align}
with onsite interaction $U$ and nearest-neighbor interaction $V$.
In the nonadiabatic limit $\om_0\to\infty$, the spinful Holstein model can be
mapped to Eq.~(\ref{eq:model-hubbard}) with $U=-4\lambda t<0$
\cite{Hirsch83a} and $V=0$. More generally, Eq.~(\ref{eq:model-hubbard}) with
$U<0$ describes a transition from a spin-gapped metallic phase at $V=0$ to a CDW insulator at $V>0$.

Finally, we investigated the spinless $t$-$V$ model
\begin{align}\label{eq:model-spinless}
  \hat{H}  
  =
  -t\sum_{i} \left( c^{\dag}_{i} c^\nag_{i+1} + \text{H.c.}\right)
  + V \sum_i \on_{i} \on_{i+1}\,
\end{align}
which for half-filling is known to have a metal-insulator
transition at $V=2t$.

We studied the above models at half-filling, corresponding to $\las
n\ras=1$ for Eqs.~(\ref{eq:model-holsteinspinful})
and~(\ref{eq:model-hubbard}), and to $\las n\ras=0.5$ for
Eqs.~(\ref{eq:model-holsteinspinless}) and~(\ref{eq:model-spinless}).
We use $t$ as the unit of energy, and set the lattice constant and $\hbar$ to one.

\section{Methods}\label{sec:method}

We simulated the models defined in Sec.~\ref{sec:model} with two different
QMC methods. First, we made use of the continuous-time interaction expansion
(CT-INT) method \cite{Rubtsov05} which  has
been successfully applied to electron-phonon lattice
models \cite{Assaad08,Hohenadler10a,Ho.As.Fe.12,HoAs12}. The
phonons are integrated out analytically, and the resulting
fermionic model with nonlocal (\ie, retarded) interactions is simulated \cite{Assaad07}. We refer to previous
publications \cite{Assaad07,Assaad08,Hohenadler10a,Ho.As.Fe.12,HoAs12} and 
reviews \cite{Gull_rev,Pavarini:155829} for technical details. 

Second, we used the stochastic series expansion (SSE) representation
\cite{Sandvik02}, which in principle provides a more favorable linear (as
compared to cubic for the CT-INT method) scaling of computer time with system
size $L$ and inverse temperature $\beta=1/\kB T$. It was previously applied
to the Holstein-Hubbard model \cite{ClHa05,hardikar:245103}. For the Holstein model,
the phonons have been treated explicitly in the occupation number basis. In order
to avoid negative vertex weights, a cut-off on the maximum phonon occupation
needs to be imposed. However, unlike exact diagonalization or DMRG methods,
the computational effort scales only linearly in the phonon cut-off, allowing us to
choose it sufficiently large to make the resulting systematic errors
completely negligible. While the employed update scheme is inherently
grand-canonical, measurements were restricted to half-filled configurations
\footnote{To avoid a sign problem in the SSE representation, instead of the particle-hole symmetric form of the interaction in
  Eq.~(\ref{eq:model-holsteinspinful}) for which $\mu=0$ corresponds to
  half-filling, we used $- g \sum_{i} \hat{Q}_{i} \hat{n}_{i}$ and a
  chemical potential $\mu=4\lambda t$}.

In the SSE representation, the directed loops algorithm \cite{Sandvik02}
allows us to update the purely fermionic models very efficiently and to reach
significantly larger system sizes compared to the CT-INT method. In contrast,
the phononic operators of the Holstein model are updated by exchanging pairs
of phonon creation and annihilation operators on the same site with diagonal
electronic operators, and vice versa, with Metropolis probabilities
\cite{hardikar:245103,SandvikPhononsa}. The number of diagonal phonon
operators between the pair enters the acceptance probabilities as an
exponent. Thus, in the nonadiabatic regime, $\omega_0\gg t$, the acceptance
rates for phononic updates are exponentially suppressed, leading to prolonged
autocorrelation times. This can be remedied to some extent by the use of
quantum parallel tempering \cite{PhysRevB.65.155113}. Still, for
$\omega_0/t=5$ we found that the CT-INT method is in fact competitive.

We measured the real-space correlation functions in the charge, pairing, and
spin channels, 
\begin{align}\nonumber\label{eq:correlctqmc}
  S_\rho(r)   &=  \las (\on_r -n) (\on_0-n) \ras\,,\\\nonumber
  S_\pi(r)   &= \las \hat{\pi}^\dag_r \hat{\pi}^\nag_0 \ras\,,\\
  S^{\alpha}_\sigma(r) &= \las \hat{S}^\alpha_r \hat{S}^\alpha_0 \ras\,.
\end{align}
Here $\alpha=x,z$, and the pairing operator is given by $\hat{\pi}^\dagger_r =
c^\dag_{r\UP} c^\dag_{r\DO}$ in the spinful case, and by $\pi^\dag_r =
c^\dag_{r} c^\dag_{r+1}$ in the spinless case. We also consider the
corresponding structure factors obtained via Fourier transformation,
\begin{equation}\label{eq:strucfact}
  S^{(\alpha)}_\nu (q) = \sum_r e^{\rmi q r} S^{(\alpha)}_\nu(r)\,.
\end{equation}

\section{Holstein models}\label{sec:holstein}

\subsection{Overview of existing work}\label{sec:overview}

We begin with an overview of important previous results for the half-filled Holstein
model.  The reason for focusing on the pure Holstein model is that it features only one phase transition, whereas
the Holstein-Hubbard model exhibits Mott and Peierls transitions
\cite{hardikar:245103,0295-5075-84-5-57001,PhysRevB87.075149}.  

\subsubsection{Numerical results}

A variety of numerical methods have been applied to the half-filled Holstein
model, most notably exact diagonalization
\cite{Fe.We.We.Go.Bu.Bi.2002,Fe.Ka.Se.We.2003,FeWeHaWeBi03} and QMC methods
\cite{Hirsch83a,ClHa05,hardikar:245103,PhysRevB.84.165123,PhysRevB87.075149},
as well as the DMRG \cite{JeZhWh99,0295-5075-84-5-57001,PhysRevB.76.155114,1742-6596-200-1-012031}. 
The first study was carried out using a QMC algorithm
\cite{Hirsch83a}. Based on results for the real-space charge correlation
function, and reassured by approximate analytical arguments, it was claimed
that the spinful Holstein model at half-filling is a Peierls insulator for
any finite phonon frequency $\om_0<\infty$ \cite{Hirsch83a}. We will show in Sec.~\ref{sec:metallicphase} that this
conclusion was incorrect.  Accordingly, all ensuing numerical works
suggest that the model instead has an extended metallic phase where
quantum lattice fluctuations destroy the dimerized
Peierls state below a critical value of the electron-phonon
coupling \cite{hardikar:245103,0295-5075-84-5-57001}.
The numerical phase
diagrams \cite{0295-5075-84-5-57001,hardikar:245103} agree with respect to the
overall features but, as discussed below, there are
nonnegligible differences regarding the phase boundaries. Despite these
quantitative uncertainties, the DMRG and QMC results rather firmly establish
the existence of a phase transition from a spin-gapped metallic phase to a
charge-ordered insulating phase with increasing electron-phonon coupling. 

Interest in the Holstein(-Hubbard) model revived when QMC simulations \cite{hardikar:245103}
suggested the existence of a metallic phase with dominant pairing
correlations, as indicated by a Luttinger parameter $\Kr>1$.
However, it was soon shown that for $\lambda>0$ charge correlations always
decay slower than pairing correlations \cite{tam:161103}.
This conflict between the numerical values of $\Kr$ and the behavior of the
correlation functions has not been resolved. An
important property of the metallic phase is the existence of a spin gap,
first pointed out in Ref.~\onlinecite{hardikar:245103}, which
arises from the pairing of electrons into spin singlets. In the language of
bosonization, the spin gap is caused by attractive backscattering of electrons,
similar to the attractive Hubbard model. A detailed discussion of the spin
gap can be found in Ref.~\onlinecite{PhysRevB87.075149}. 
The most compelling
evidence for a spin gap is the dominance of charge correlations over spin
correlations, which is not possible in a gapless Luttinger liquid \cite{Voit98} but 
observed in the Holstein model in the entire metallic
phase \cite{PhysRevB87.075149}, as well as the numerical finding of a
Luttinger parameter $K_\sigma<1$ incompatible with the SU(2) spin symmetry
of the Holstein model \cite{PhysRevB87.075149} (see Sec.~\ref{sec:correlations}).
An important corollary of the existence of a spin gap in the metallic phase
is that the relevant fixed point in the thermodynamic limit is not the Luttinger
liquid but the Luther-Emery liquid, which is much more
difficult to analyze.

\subsubsection{Analytical results}

In the static (classical phonon or adiabatic) limit $\om_0\to0$ a mean-field ansatz 
$Q_i=(-1)^i \Delta$ reveals the Peierls instability inherent to
1D systems without quantum fluctuations.
Because the energy gain outweighs the cost for dimerization, any nonzero
electron-phonon coupling leads to an insulating Peierls state. This
conclusion holds for both the spinful and the spinless Holstein model.
In the opposite, anti-adiabatic limit $\om_0\to\infty$, the Holstein model
can be mapped to the attractive Hubbard model
with an instantaneous interaction $U=-4\lambda t$ \cite{Hirsch83a}. The half-filled attractive
Hubbard model is metallic for any $U$, and has a spin gap $\Delta_\sigma\sim
e^{-v_\text{F}/\mathcal{U}}$ where $\mathcal{U}$ is the effective
backscattering matrix element \cite{Giamarchi}.  While representing valuable
limiting cases, neither the static ($\om_0\to0$) nor the instantaneous ($\om_0\to\infty$)
limit captures the theoretical problem of electrons coupled to quantum phonons.

The strong-coupling approximation presented by Hirsch and Fradkin \cite{Hirsch83a}
starts from electrons paired into spin singlets and considers the effect of
second-order hopping processes to derive an effective hardcore boson
model with pair hopping $\tilde{t}$ and nearest-neighbor repulsion
$\tilde{V}$, both functions of $\lambda$ and $\om_0$. The strong-coupling results give $\tilde{V}/2\tilde{t}>1$ for any
$\om_0<\infty$ which, according to the bosonization results for hardcore
bosons, implies that the system is in an insulating, charge-ordered phase for
any $\lambda>0$. In contrast, a similar approximation
for the spinless Holstein model gives $\lambda_c>0$. [Interestingly, the same difference between the spinful and
spinless cases is also predicted for the Su-Schrieffer-Heeger model. In
contrast to the Holstein model, these analytical predictions ($\lambda_c>0$
for the spinless model, $\lambda_c=0$ for the spinful model) have been
confirmed by numerical simulations, see Ref.~\onlinecite{PhysRevB.91.245147} and
references therein.] The strong-coupling approximation highlights the relation
of the spinful Holstein model to hardcore bosons (corresponding to electron
pairs), but is considered to be unreliable in the weak-coupling regime.
In particular, for sufficiently small $\lambda$, the approximation of electron pairs as hardcore,
onsite objects breaks down. Instead, the system may be regarded as consisting
of interacting singlet bipolarons whose size depends on $\lambda$ and $\omega_0$.

Because we are considering a 1D system, important insights can be gained from the
bosonization method together with the RG. For electron-phonon models, such an
approach involves an additional approximation. The RG flow is carried out in two steps:
From high energies down to the energy for phonon excitations, then (assuming
an instantaneous interaction) from the phonon energy down to
zero \cite{PhysRevB.71.205113,Ba.Bo.07}. Additionally, the  momentum dependence of the phonon-mediated
electron-electron interaction is
usually neglected. For the spinful Holstein model, the RG
suggests $\lambda_c=0$ for $\omega_0<\infty$, whereas $\lambda_c>0$ is found for
the spinless Holstein model  \cite{PhysRevB.71.205113,Ba.Bo.07}. The results for the
spinful Holstein model are often quoted as evidence for the
absence of a metallic phase, despite the above-mentioned
limitations. An extended metallic phase was recently confirmed 
using the functional RG method \cite{Barkim2015}, which takes into account
the frequency dependence of the interaction.

\subsection{Existence of a metallic phase}\label{sec:metallicphase}

\begin{figure}[t]
  \includegraphics[width=0.5\textwidth]{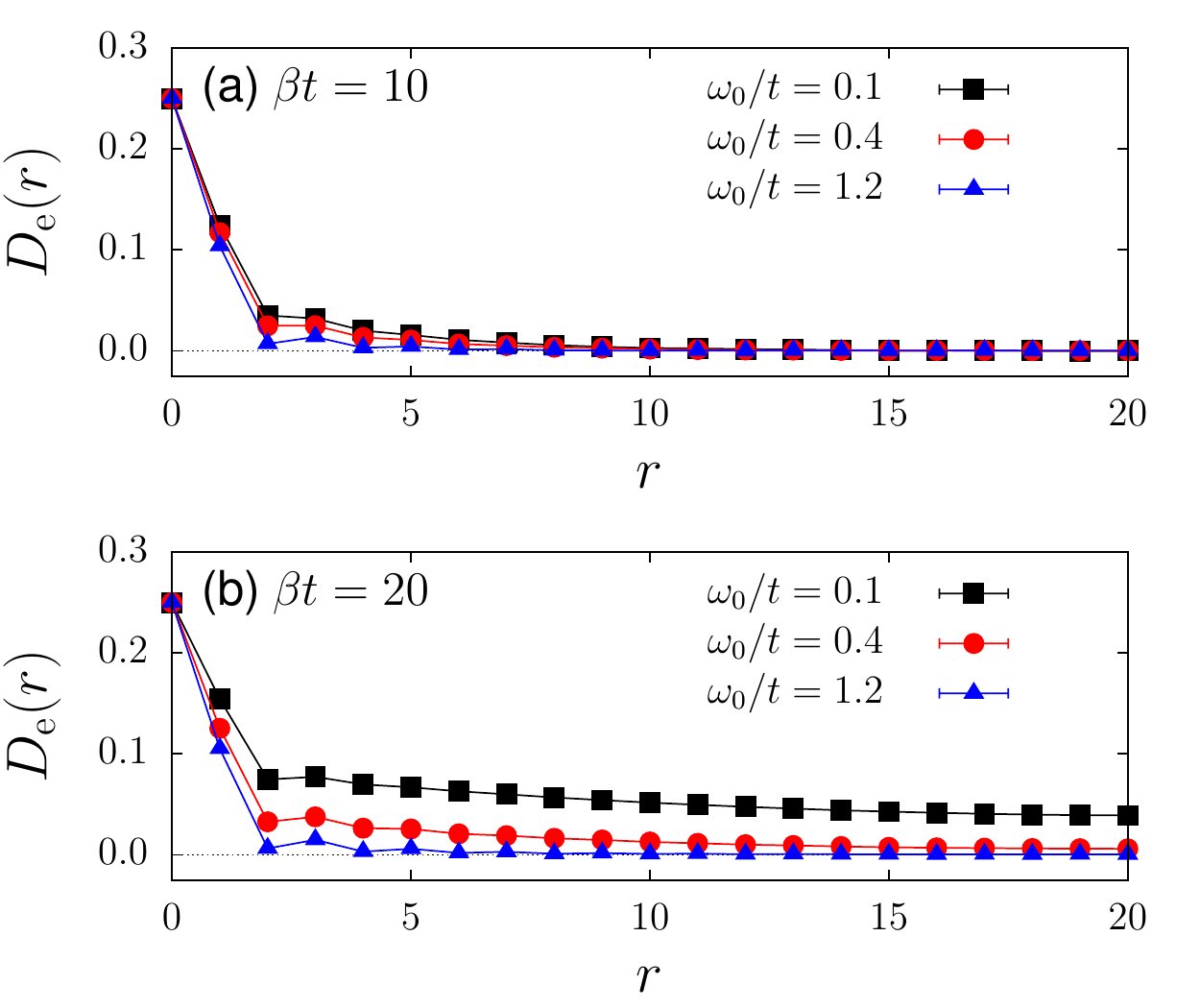}
  \caption{\label{fig:hirschfradkinspinless} (Color online) 
    Charge correlations [Eq.~(\ref{eq:De})] for the spinless
    Holstein model at (a) $\beta t=10$, (b) $\beta t=20$.  Here,
    $\lambda=0.81$ and $L=42$. The results can be compared to Fig.~7(b)
    in Ref.~\onlinecite{Hirsch83a}. Results obtained with the CT-INT
    method. Here and in subsequent figures, the error bars are
    typically smaller than the symbols.
}
\end{figure}

To settle the argument, we
revisit the QMC results of Hirsch and Fradkin (HF)  \cite{Hirsch83a}, the only
numerical results that suggest the absence of a metallic phase.  Their
conclusions rely on data for the real-space charge and lattice
correlation functions [Figs.~7(b) and~8(b) in
Ref.~\onlinecite{Hirsch83a}]. For the parameters considered, the data suggest
long-range order (up to the system size considered) for charge and lattice
correlations  in the spinful Holstein model both at low and high phonon
frequencies, from which the authors conclude the absence of a metallic
phase. In contrast, a crossover from long-range to short-range order was
observed in the spinless model upon increasing the phonon frequency, which is
compatible with a Peierls transition as a function of $\om_0$ or,
equivalently, $\lambda$.

The coupling constant used in Ref.~\onlinecite{Hirsch83a}, denoted here as
$\lambda_\text{HF}$, is related to $\lambda$ via
$\lambda=\lambda_\text{HF}^2/(4Kt)$. HF set $t=1$ and  $K=0.25$, so that
$\lambda=\lambda_\text{HF}^2$. They considered $\lambda_\text{HF}=0.9$
[$\lambda=0.81$] for the spinless model, and $\lambda_\text{HF}=0.9/\sqrt{2}$
[$\lambda=0.405$] for the spinful model, as well as phonon frequencies
$\om_0/t=0.4$ and $\om_0/t=1.2$. The inverse temperature was $\beta t
= 10$ for $\om_0/t=0.4$ and $\beta t = 5$ for $\om_0/t=1.2$. The system size
was $L=40$ sites. In contrast to the continuous-time methods used here, the
results of Ref.~\onlinecite{Hirsch83a} also have a systematic Trotter error.

Using the CT-INT method, we calculated the charge correlation function
considered by HF \cite{Hirsch83a},
\begin{equation}\label{eq:De}
  D_\text{e}(r)
  =
  \frac{1}{L}\sum_{i\sigma\sigma'} (-1)^r
  \left[
    \las \on_{i\sigma} \on_{i+r,\sigma'}\ras
    -
    n^2
  \right]\,,
\end{equation}
which is equivalent to $(-1)^r S_\rho(r)$, cf. Eq.~(\ref{eq:correlctqmc}).

Results for $D_\text{e}(r)$ of the spinless model are shown in
Fig.~\ref{fig:hirschfradkinspinless} for $L=42$. For the lowest temperature used by HF, $\beta t=10$,
Fig.~\ref{fig:hirschfradkinspinless}(a) reveals the absence of long-range
charge correlations for all values of $\om_0$ considered (in conflict with
Ref.~\onlinecite{Hirsch83a}). At a lower temperature, $\beta t = 20$, we find
results similar to those of HF, namely long-range order for $\om_0/t=0.4$ but not for $\om_0/t=1.2$. We attribute the
discrepancies for $\beta t=10$ to
autocorrelations, which can be significant for the parameters considered \cite{PhysRevB.71.245111},
thereby giving rise to spuriously enhanced charge
correlations. Autocorrelations are significantly smaller and properly
accounted for in the present results. The results in
Fig.~\ref{fig:hirschfradkinspinless}(b) are consistent with a Peierls
transition as a function of phonon frequency in the {\it spinless} Holstein
model, as claimed in Ref.~\onlinecite{Hirsch83a}.

\begin{figure}[t]
  \includegraphics[width=0.5\textwidth]{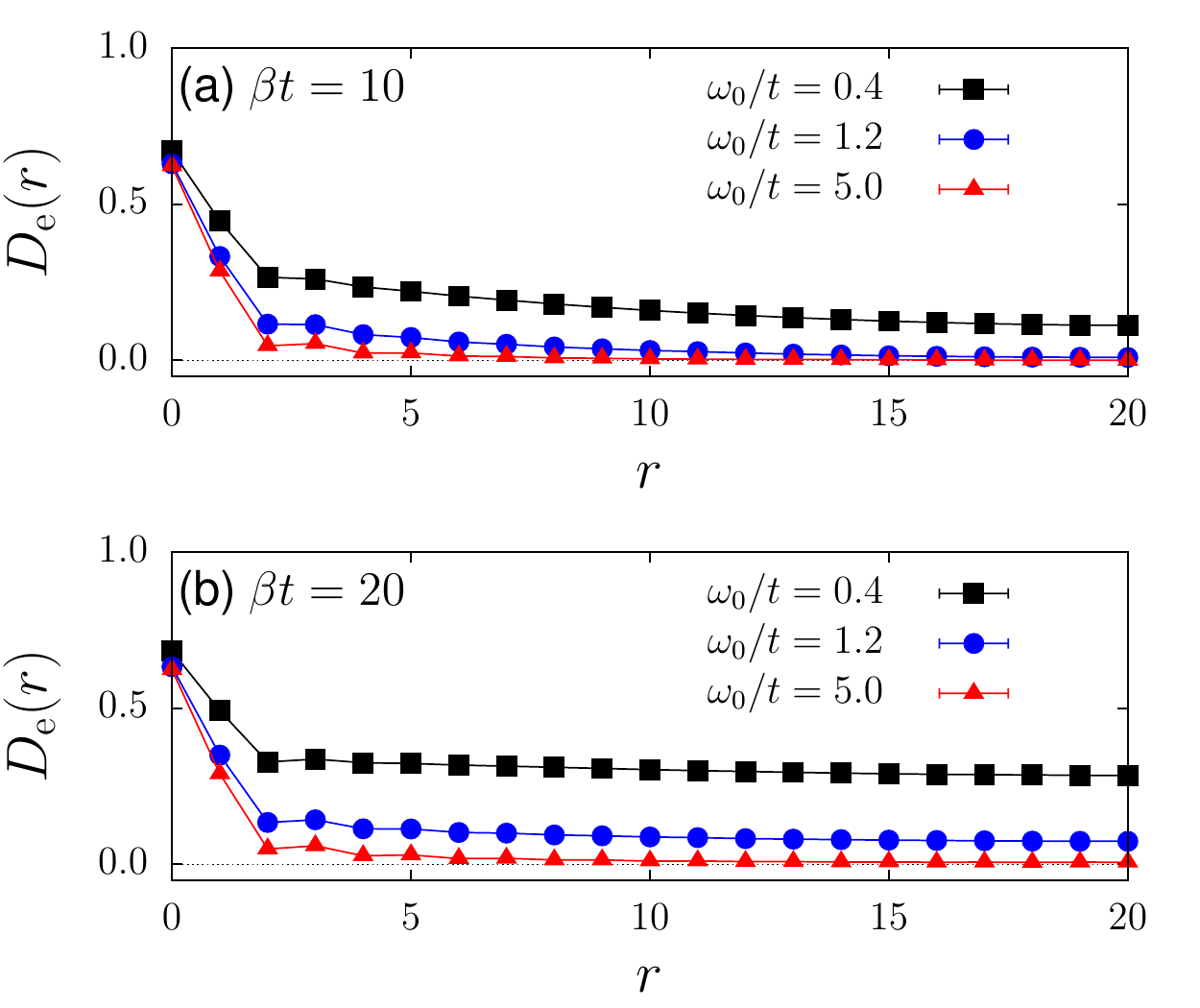}   
  \caption{\label{fig:hirschfradkinspinful} (Color online) 
    Charge correlations [Eq.~(\ref{eq:De})]
    for the spinful Holstein model at (a) $\beta t=10$, (b) $\beta t=20$. 
    Here, $\lambda=0.405$ and $L=42$.
    The results can be compared to Fig.~8(b) in Ref.~\onlinecite{Hirsch83a}.
    Results obtained with the CT-INT method.
}
\end{figure}

Results for the {\it spinful} Holstein model are shown in
Fig.~\ref{fig:hirschfradkinspinful}. For parameters comparable to those used by HF,
Fig.~\ref{fig:hirschfradkinspinful}(a), we find long-range correlations up to
the maximal distance for $\om_0/t=0.4$, but not for $\om_0/t=1.2$. However,
both cases show significantly weaker charge correlations than those in
Ref.~\onlinecite{Hirsch83a}, a fact that we again attribute to autocorrelations.
At a lower temperature, shown in Fig.~\ref{fig:hirschfradkinspinful}(b),
similar to the spinless case, we find results which look similar to those of
HF, namely long-range charge order for both $\om_0/t=0.4$ and $\om_0/t=1.2$.
However, if we increase $\om_0$ further, we again find a transition to a state
without long-range order, consistent with a metallic state
at $\lambda>0$. 

\begin{figure}[t]
  \includegraphics[width=0.5\textwidth]{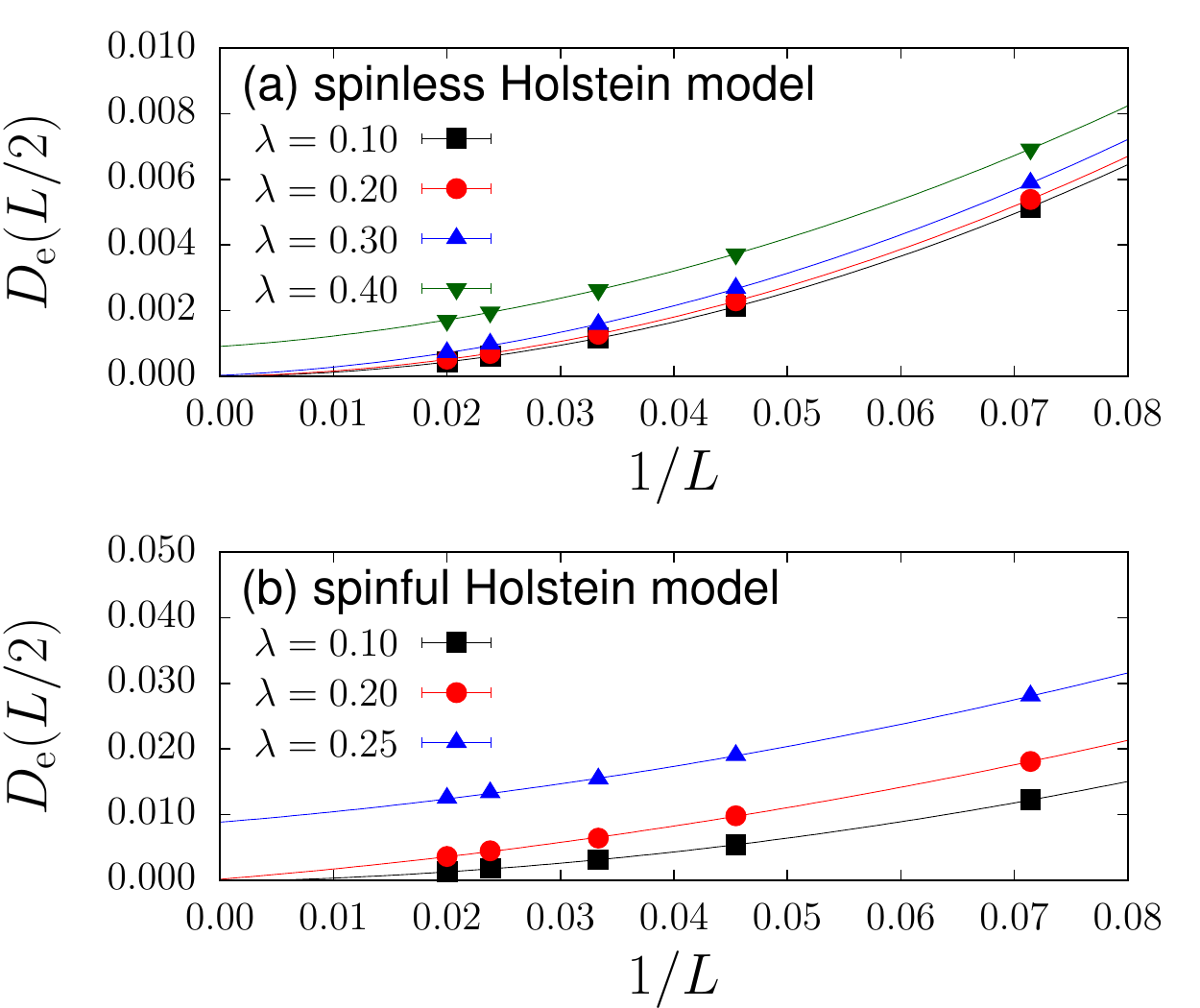}
  \caption{\label{fig:longrangeorder} (Color online) 
  Finite-size scaling of the charge correlations at the largest distance
  $r=L/2$. Lines are fits to second-order polynomials.  Here, $\beta t=L$, 
  (a) $\om_0/t=0.1$, (b) $\om_0/t=0.5$. Results obtained with the CT-INT method.
}
\end{figure}

To estimate $\lambda_c$, we carried
out a finite-size extrapolation of the charge correlations at the largest
distance. The results in Fig.~\ref{fig:longrangeorder}
are consistent with extended metallic  regions in both the
spinless and the spinful model. We have verified that our low-temperature results are
representative of the ground state. The onset of long-range order in
Fig.~\ref{fig:longrangeorder} is consistent with the best available
estimates for the critical values of the spinless ($\lambda_c\approx 0.39$ \cite{Hohenadler06}) and
the spinful Holstein model ($\lambda_c\approx 0.23$
\cite{ClHa05,hardikar:245103,0295-5075-84-5-57001,1742-6596-200-1-012031}). 

\subsection{Real-space correlation functions}\label{sec:correlations}

In 1D metals, the decay of correlation functions is expected to be 
parametrized by the charge and spin Luttinger parameters $\Kr$ and
$\Ks$, respectively.  However, previous numerical and
analytical work revealed a number of difficulties in applying Luttinger liquid theory to the spinful Holstein model, whereas good
agreement was observed for the spinless Holstein model \cite{Ej.Fe.09}. Most
importantly, the values of $\Kr$ obtained with exact numerical methods are incompatible with the
numerically determined correlation functions. 

\subsubsection{Spinless Holstein model}

To set the stage for the discussion of the spinful case, it is useful
to first consider the spinless Holstein model~(\ref{eq:model-holsteinspinless}).
The bosonization method applied to 1D spinless fermions gives the following 
results for the real-space correlation functions (we use $x$ to
denote distances in the continuum limit, and $r$ for distances on a lattice) \cite{Voit94}:
\begin{align}\nonumber\label{eq:correl:LLspinless}
  S_\rho(x) 
  &=
  -
  \frac{\Kr}{2\pi^2x^2} + \frac{A_\rho}{x^{2\Kr}}\cos(2\kF x)\,,
  \\
  S_\pi(x) 
  &=
  \phantom{-}
  \frac{A_\pi}{x^{2\Kr^{-1}}} \,.
\end{align}
Here, $\rho$ ($\pi$) denotes the charge (pairing) sector. 

According to these results, the exponents determining the power-law decay of
correlations depend only on $\Kr$.
Equation~(\ref{eq:correl:LLspinless}) further implies that $q=0$ pairing
correlations dominate (\ie, decay slowest) in the case of attractive
interactions ($K_\rho>1$), whereas $q=2\kF$ charge correlations dominate in
the case of repulsive interactions ($K_\rho<1$). For noninteracting
electrons ($\Kr=1$), charge and pairing correlations both decay as $1/x^2$.

\begin{figure}[ht]
  \includegraphics[width=0.5\textwidth]{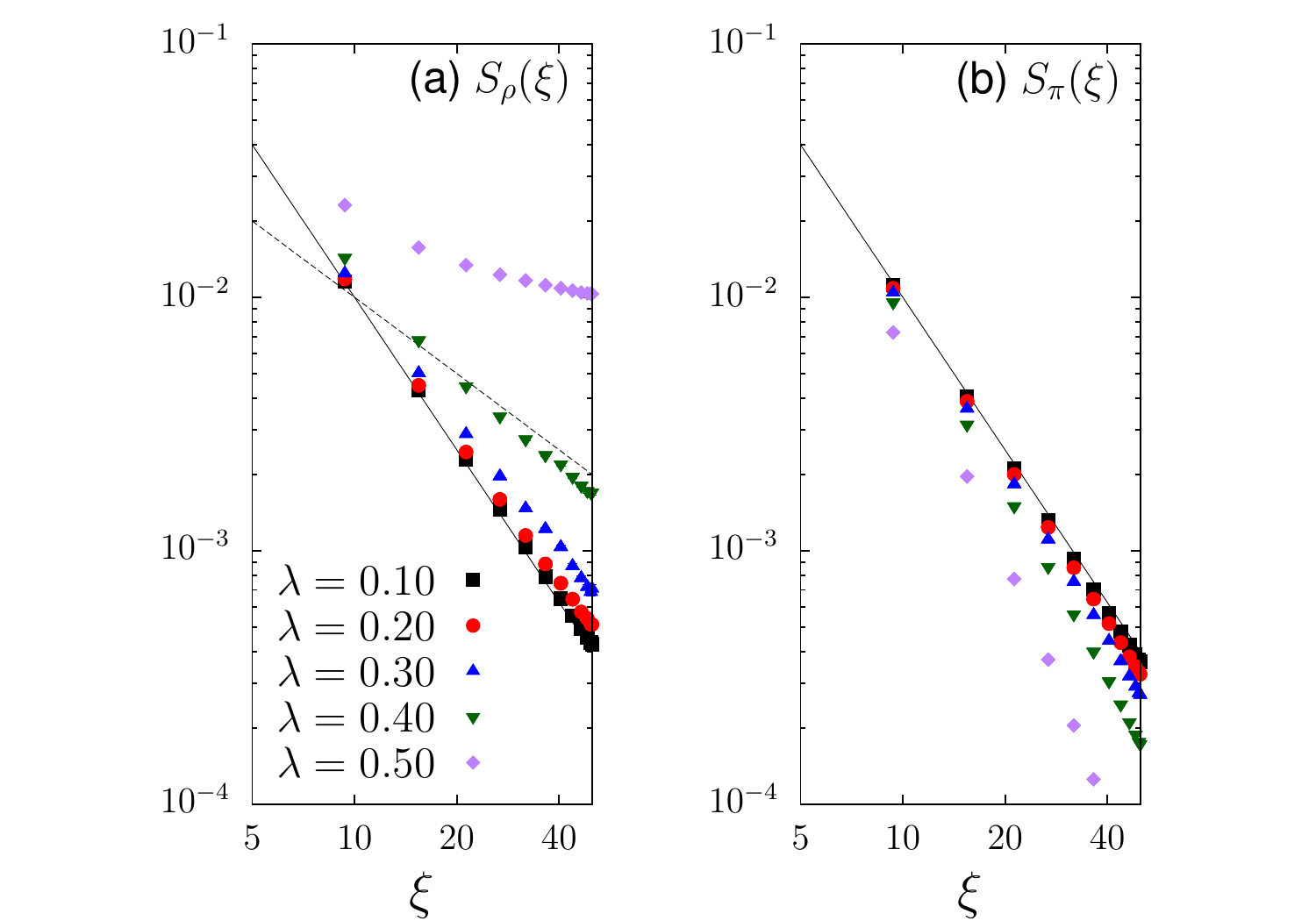}
  \caption{\label{fig:correlspinlessdifflambda} (Color online) 
    Charge and pairing correlation functions of the spinless Holstein
    model.    Solid (dashed) lines illustrate $c/x^2$ ($c/x$).
     Here, $\om_0/t=0.1$, $L=\beta t=50$.
    Results obtained with the CT-INT method.
}
\end{figure}

\begin{figure}[ht]
  \includegraphics[width=0.5\textwidth]{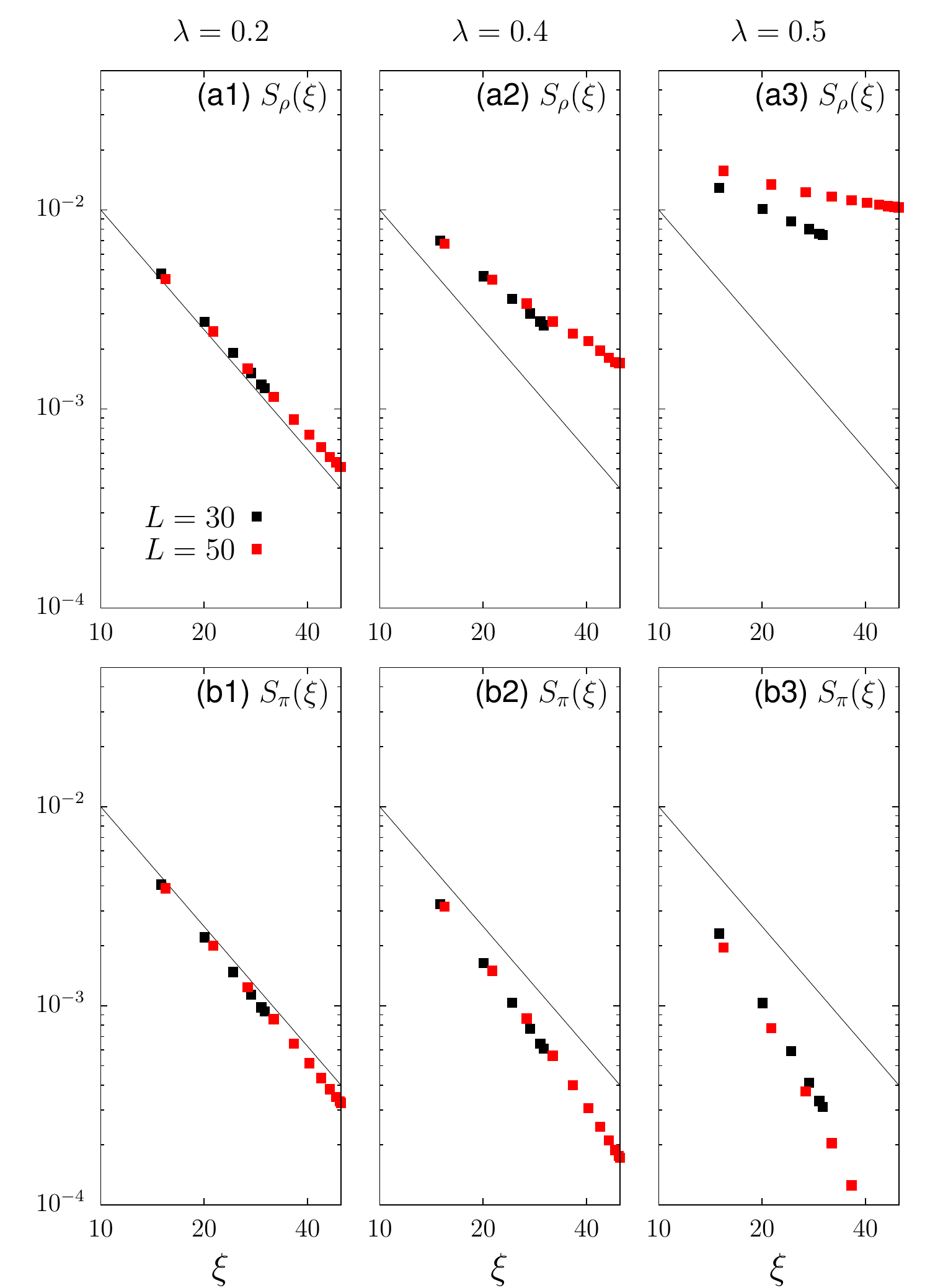}
  \caption{\label{fig:powerlawspinless} (Color online) 
    Charge and pairing correlations of the spinless Holstein
    model. Solid lines illustrate $c/x^2$.
    Here, $\om_0/t=0.1$,
    $\beta t=L$. Results obtained with the CT-INT method. 
}
\end{figure}

We calculated the correlation functions~(\ref{eq:correlctqmc}) with the CT-INT
method, and plot the results in terms of the conformal distance
\cite{Cardybook}  $\xi = L\sin ({\pi r}/{L})$ to remove effects of the
periodic boundary conditions.
The results are shown in Fig.~\ref{fig:correlspinlessdifflambda}. For
$\lambda<\lambda_\text{c}=0.39(2)$, the numerical data are consistent with a
power-law decay, as suggested by Eq.~(\ref{eq:correl:LLspinless}). Because
charge correlations dominate over pairing correlations, we conclude that
$K_\rho<1$. Moreover, Fig.~\ref{fig:correlspinlessdifflambda} suggests that
$K_\rho$ decreases from the noninteracting value $K_\rho=1$ with increasing
$\lambda$, in accordance with DMRG results \cite{Ej.Fe.09}. The
critical point is expected at $K_\rho=1/2$ (see discussion in
Sec.~\ref{sec:discussion}), suggesting a $1/x$ decay of
charge correlations at $\lambda_c$ that is in satisfactory agreement with the data.

Figure~\ref{fig:powerlawspinless} provides a more stringent test of the
relationship between our numerical data and the bosonization results in
Eq.~(\ref{eq:correl:LLspinless}). It shows results for the correlation
functions for different system sizes $L$ at a given value of $\lambda$.
If the correlations decay with a power-law 
determined by $K_\rho$, we expect data for
different $L$ to fall onto the same straight line in a log-log
plot (\ie, to have the same exponent). According to
Fig.~\ref{fig:powerlawspinless}, this is indeed the case for
$\lambda=0.2$. In contrast, for $\lambda=0.4$ and $\lambda=0.5$, corresponding
to the Peierls phase, we find a violation of this ``scaling''. Instead of a
power-law decay, the charge correlations exhibit long-range order, and the
corresponding pairing correlations [Fig.~\ref{fig:powerlawspinless}(b3)] are
consistent with an exponential decay at large distances.

\begin{figure}
  \includegraphics[width=0.5\textwidth]{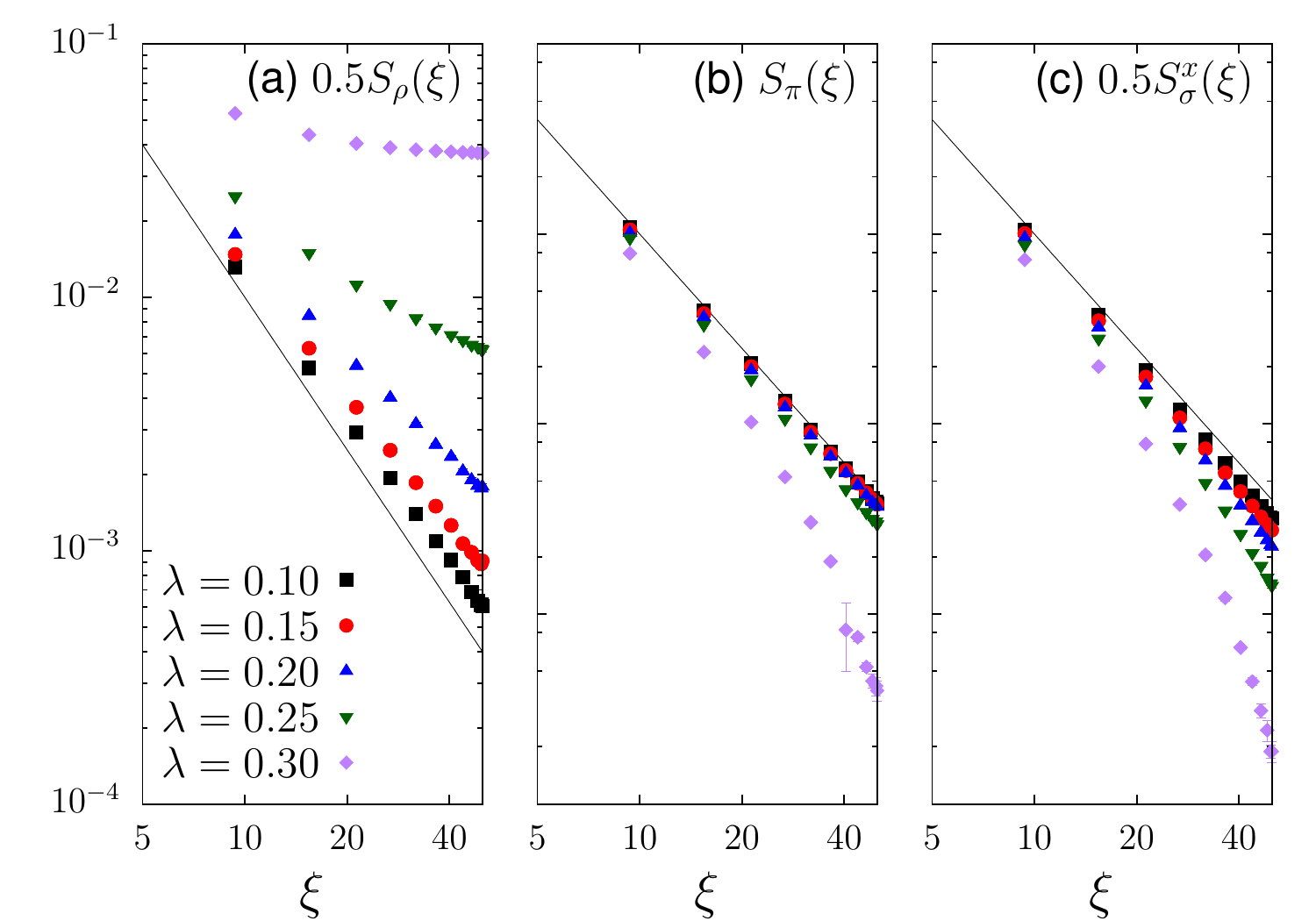}
  \caption{\label{fig:correlspinfuldifflambda} (Color online) 
    Charge, pairing, and spin correlation functions of the spinful Holstein
    model. The solid line illustrates $c/x^2$.
    Here, $\om_0/t=0.5$, $\beta t = L = 50$.    
    Results obtained with the CT-INT method.
}
\end{figure}

\subsubsection{Spinful Holstein model}

For spinful fermions, the bosonization gives \cite{PhysRevB.39.4620,Voit94}
\begin{align}\nonumber\label{eq:correl:LL}
  S_\rho(x) 
  &=
  -\frac{\Kr}{\pi^2 x^2} + \frac{B_\rho}{x^{\Kr+\Ks}}\cos(2\kF x)\,,
  \\\nonumber
  S_\pi(x) 
  &=
  \phantom{-}
  \frac{B_\pi}{x^{\Kr^{-1}+\Ks}}\,, 
  \\
  S_{\sigma}(x)
  &=  
  -\frac{\Ks}{4\pi^2 x^2} + \frac{B_\sigma}{x^{\Kr+\Ks}}\cos(2\kF x) 
\end{align}
for a Luttinger liquid without a spin gap, and \cite{Voit94}
\begin{align}\nonumber\label{eq:correl:LE}
  S_\rho(x) 
  &=
  \frac{C_\rho}{x^2} + \frac{C'_\rho}{x^{\Kr}}\cos(2\kF x)\,,
  \\
  S_\pi(x) 
  &=
  \frac{C_\pi}{x^{\Kr^{-1}}} 
\end{align}
for a Luther-Emery liquid with a gap for spin excitations. 
Given SU(2)
spin symmetry, the value of $\Ks$ in Eq.~(\ref{eq:correl:LL}) is fixed to 1,
while the Luther-Emery correlation functions [Eq.~(\ref{eq:correl:LE})] are 
obtained by setting $\Ks=0$. The Peierls phase with a spin gap and long-range
charge correlations formally corresponds to $\Kr=\Ks=0$.
In Eqs.~(\ref{eq:correl:LL}) and~(\ref{eq:correl:LE}), we included only the
dominant $q=0$ part for the pairing correlators. Moreover, in
Eq.~(\ref{eq:correl:LL}), we neglected the $q=4\kF$ charge term because it is always
subdominant for the cases considered here. Equations~(\ref{eq:correl:LL})--(\ref{eq:correl:LE}) 
ignore possible logarithmic corrections, which can arise from
marginally irrelevant operators \cite{0305-4470-22-5-015,PhysRevB.39.4620}.

\begin{figure}
  \includegraphics[width=0.5\textwidth]{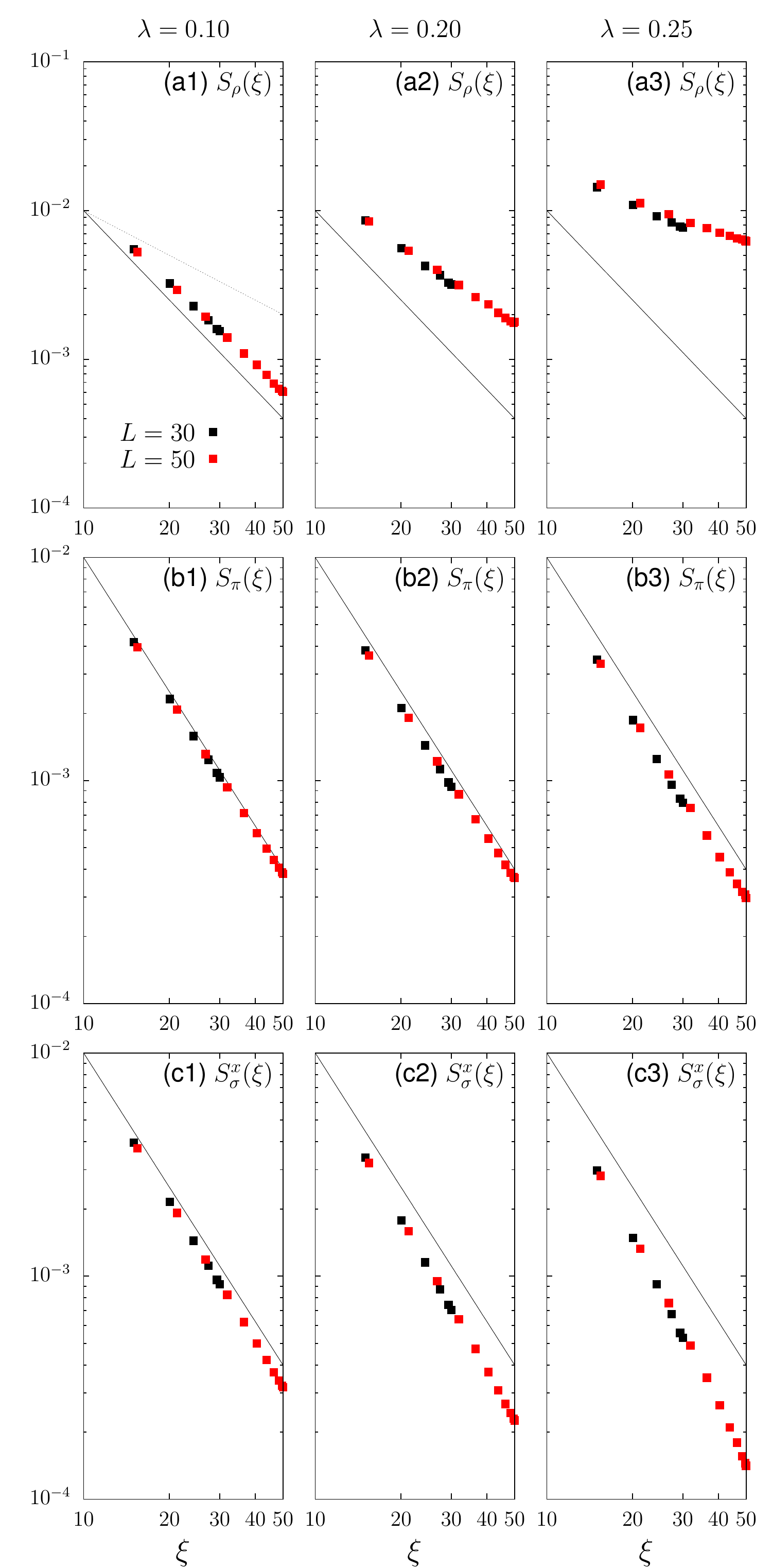}
  \caption{\label{fig:powerlawspinful} (Color online) 
    Charge, pairing, and spin correlation functions of the spinful Holstein
    model. Solid (dashed) lines illustrate $c/x^2$ ($c/x$).
    Here, $\om_0/t=0.5$, $\beta
    t=L$. Results obtained with the CT-INT method.
}
\end{figure}

In both Luttinger liquids and Luther-Emery liquids, charge correlations
dominate over pairing correlations for $K_\rho<1$. However, for Luttinger
liquids $2\kF$ charge and spin correlations have
exactly the same exponent [Eq.~(\ref{eq:correl:LL})]. This degeneracy can be
lifted in favor of dominant charge correlations (and exponential spin
correlations) by a spin gap \cite{Voit98}. 
As argued before \cite{PhysRevB87.075149,tam:161103,PhysRevB.84.165123}, numerical
results for the correlation functions of the
spinful Holstein model are consistent with a spin
gap. Figure~\ref{fig:correlspinfuldifflambda} shows results for different values of $\lambda$. Starting with
a $1/x^2$ decay for $\lambda=0$, we observe an enhancement of charge
correlations with increasing $\lambda$, and a suppression of both pairing and
spin correlations. This behavior is captured by the Luther-Emery
correlation functions in Eq.~(\ref{eq:correl:LE}) with $K_\rho<1$, but not by the Luttinger liquid
expressions of Eq.~(\ref{eq:correl:LL}). The dominance of charge over pairing
correlations
\cite{PhysRevB87.075149,tam:161103,PhysRevB.84.165123,TeArAo05,PhysRevB.76.155114}
contradicts earlier claims of a metallic phase with dominant superconducting
correlations (\ie, $K_\rho>1$) \cite{ClHa05}, and the claim of $K_\rho=1$ in
the metallic phase \cite{hardikar:245103}. 

Figure~\ref{fig:powerlawspinful} shows charge, pairing and spin
correlation functions for different system sizes, with $\lambda$ fixed for
each column. As for the spinless model, we can test if the data exhibit the
scaling expected for power-law correlations. For $\lambda=0.1$, deep in the
metallic phase, the results in Figs.~\ref{fig:powerlawspinful}(a1), (b1), and
(c1) are consistent with a power-law decay. The absence of a clear exponential decay in
Fig.~\ref{fig:powerlawspinful}(c1) suggests that the system sizes are not
sufficient to reach the Luther-Emery fixed point where
Eq.~(\ref{eq:correl:LE}) holds.  A
clear exponential decay of spin correlations in the metallic phase has been
observed in DMRG studies carried out at higher phonon frequencies where the
metallic phase extends to larger values of $\lambda$ \cite{TeArAo05,PhysRevB.76.155114}.
The crossover from Luttinger liquid to Luther-Emery liquid behavior as a
function of distance will be illustrated for the numerically more accessible
attractive Hubbard model in Sec.~\ref{sec:attractivehubbard}.

For a stronger coupling $\lambda=0.2$
[Figs.~\ref{fig:powerlawspinful}(a2), (b2), (c2)], close to the critical point,
we observe signatures of long-range charge order, and of a break-down of
power-law scaling.  Finally, for $\lambda=0.25$
[Figs.~\ref{fig:powerlawspinful}(a3), (b3), (c3)], corresponding to the Peierls
phase, deviations from scaling are visible in all three channels. Charge
correlations clearly reflect the long-range order, while pairing and spin
correlations are consistent with an exponential decay.

\subsubsection{Luttinger liquid parameter $\Kr$}\label{sec:Krho}

Our numerical results for the correlation functions may be explained using
the bosonization expressions with suitable exponents. It is therefore of
particular interest to calculate $K_\rho$ as a function of $\om_0$ and $\lambda$.

$\Kr$ is routinely extracted from numerical results for the charge structure factor, 
\begin{equation}\label{eq:krhofit}
  \Kr
  =
  s \pi \lim_{q\to 0} S_\rho(q) / q\,,
\end{equation}
where $s=1$ ($s=2$) for spinful (spinless) models. The relation of $K_\rho$
to the $q=0$ charge fluctuations follows from $S_\rho(x)$ in Eqs.~(\ref{eq:correl:LLspinless})
and~(\ref{eq:correl:LL}), the first term of which is directly proportional to
$K_\rho$. Although $K_\rho$ is defined in the thermodynamic limit, a finite-size
estimate can be obtained from
\begin{equation}\label{eq:krhoL}
\Kr(L) = s \pi  S_\rho(q_1) / q_1\,,
\end{equation}
where $q_1=2\pi/L$ is the smallest nonzero wavevector; finite-size scaling
then gives, in principle, the physical value of $\Kr$, although the
scaling function is in general not known. The use of Eq.~(\ref{eq:krhoL}) is usually motivated by
simplicity and the absence of (multiplicative) logarithmic corrections to the
$q=0$ term in the correlation functions. Logarithmic corrections to the
power-law decay of correlation functions may arise from marginally irrelevant
operators \cite{0305-4470-22-5-015,PhysRevB.39.4620,Lukyanov1998533}, although we are not aware of
explicit results for corrections to the $q=0$ term. Moreover, for the
Luther-Emery fixed point, the identification of the constant $C_\rho$ with
$K_\rho$ is expected to hold only for small spin gaps. We will demonstrate below
that the spinful Holstein model, and also the attractive Hubbard model, are
examples for which the determination of $K_\rho$ from Eq.~(\ref{eq:krhoL})
is problematic.

\begin{figure}
  \includegraphics[width=0.5\textwidth]{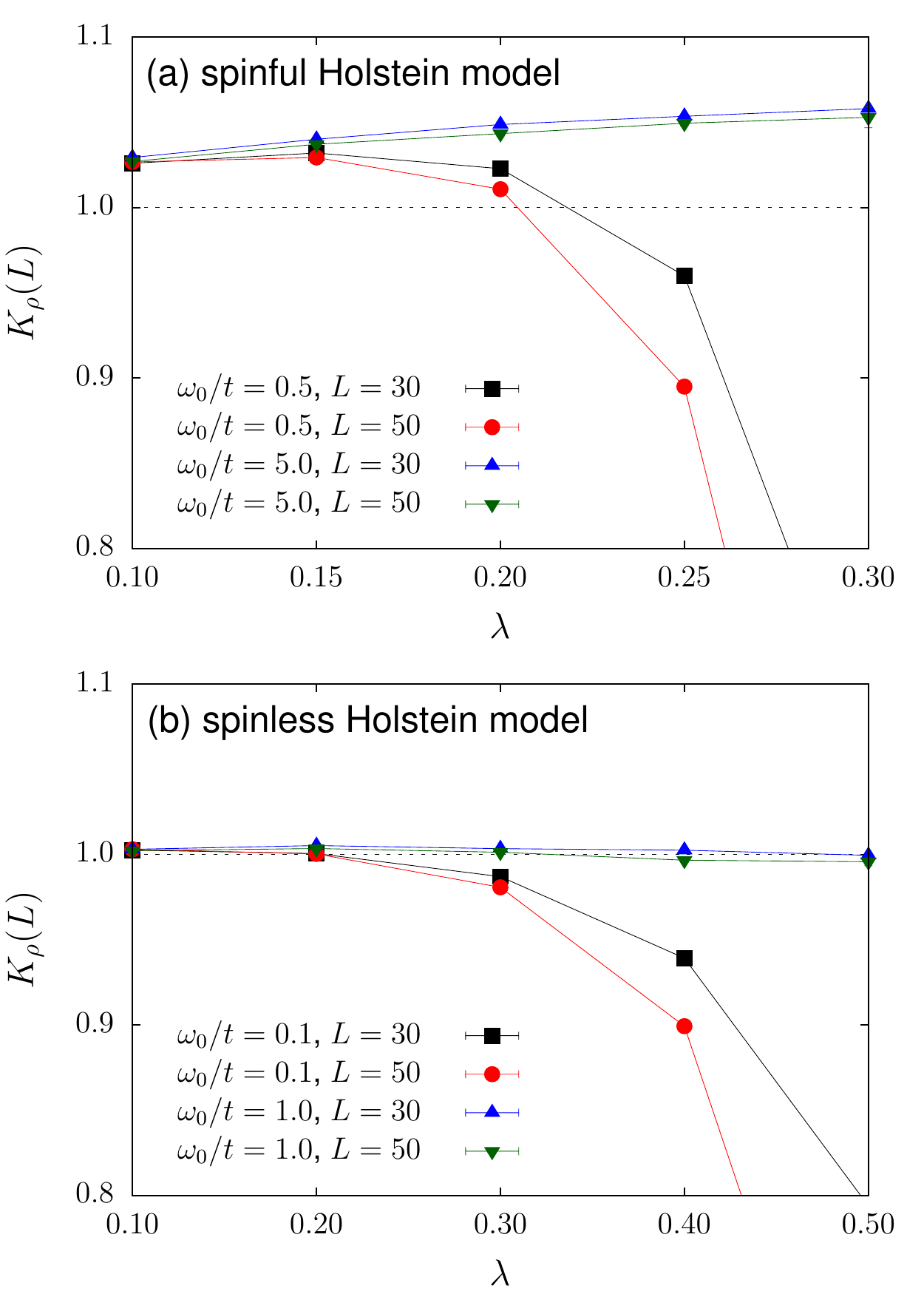}
  \caption{\label{fig:krhofromdensitylambda} (Color online) 
  $\Kr(L)$ [Eq.~(\ref{eq:krhoL})] for  (a) the spinful and (b) the spinless
  Holstein model. Here, $\beta t=L$. 
  Results obtained with the CT-INT method.
}
\end{figure}

Using Eq.~(\ref{eq:krhoL}), $K_\rho$ for the spinful Holstein and the
Holstein-Hubbard model has previously been extracted from QMC
\cite{ClHa05,hardikar:245103} and DMRG \cite{1742-6596-200-1-012031}
results. The large system sizes used (up to several hundred lattice sites)
suggest  the possibility of a reliable extrapolation to the infinite system. These
works reported $K_\rho>1$ in the metallic phase
\cite{ClHa05,1742-6596-200-1-012031}, thereby implying dominant pairing
correlations. Later, it was argued that $K_\rho>1$ due to logarithmic
corrections and that the true value is $K_\rho=1$ in the metallic phase
\cite{hardikar:245103}, as in the attractive Hubbard model. However, the
observation of dominant charge correlations (see above and
Refs.~\onlinecite{tam:161103,PhysRevB.84.165123,TeArAo05,PhysRevB.76.155114,PhysRevB87.075149})
in the metallic phase seems compatible only with $K_\rho<1$.

We used the CT-INT method to
calculate $K_\rho(L)$ via Eq.~(\ref{eq:krhoL}) for the spinful  Holstein
model. The results for $\omega_0/t=0.5$ as a function of $\lambda$ are shown  Fig.~\ref{fig:krhofromdensitylambda}.
In agreement with previous work, we find $K_\rho(L)>1$ for $\lambda>0$ in the
metallic phase for the spinful Holstein model, see Fig.~\ref{fig:krhofromdensitylambda}(a).
Starting from $\lambda=0$, $K_\rho(L)$ initially increases with
increasing $\lambda$, before it starts to decrease in the vicinity of the 
critical value $\lambda_\text{c}=0.23(1)$. A very similar behavior was observed before \cite{hardikar:245103}.
Crucially, $K_\rho(L)$ takes on even larger values
if we increase the phonon frequency from $\om_0/t=0.5$ to $\om_0/t=5$; for the
latter, the metallic phase extends up to $\lambda_\text{c}\approx0.5$
\cite{hardikar:245103,0295-5075-84-5-57001}. This observation seems to rule out
retardation effects as the reason for the slow convergence of $K_\rho$ \cite{tam:161103}.

Further insight can be gained from the spinless Holstein model. In
Fig.~\ref{fig:krhofromdensitylambda}(b), we see that $K_\rho(L)\lesssim1$ 
even for the rather small lattice sizes accessible with the CT-INT
method. In contrast to the spinful model, the finite-size extrapolated
values of $\Kr$ are always smaller than 1 for nonzero
$\lambda$ \cite{Ej.Fe.09}, consistent with the dominant charge correlations
in Fig.~\ref{fig:correlspinlessdifflambda}. Moreover,
the DMRG was able to show that $K_\rho=1/2$ coincides within the numerical
accuracy with the critical point of the
Peierls transition, as expected from bosonization
results for the related $t$-$V$ model of spinless fermions \cite{Ej.Fe.09}.

By combining the above results, we arrive at the following conclusions. For
the spinful Holstein model, the
use of Eq.~(\ref{eq:krhoL}) gives results for $K_\rho$ that are
inconsistent with the correlation functions, which show a 
generic dominance of charge over pairing correlations. Even in the absence of
accurate estimates of $K_\rho$, the behavior of the correlation functions
hence establishes the repulsive nature of the Luther-Emery phase.
In contrast,
for the spinless Holstein model, the values of $K_\rho$ determined by
large-scale DMRG calculations \cite{Ej.Fe.09} are compatible \footnote{In contrast to the spinless $t$-$V$ model
  discussed in Sec.~\ref{sec:spinlessfermions}, the values of $K_\rho$
  reported in Ref.~\onlinecite{Ej.Fe.09} do not exactly match the power-law exponents
  of the correlation functions. This suggests that the finite-size
  extrapolation of $K_\rho(L)$ is less reliable for the spinless Holstein
  model.} with the behavior of correlation functions. Since the retardation of the phonon-mediated
interaction is identical for the spinful and the spinless model for the same $\omega_0/t$
[$\omega_0/t$ is actually smaller in Fig.~\ref{fig:krhofromdensitylambda}(b)
than in Fig.~\ref{fig:krhofromdensitylambda}(a)], it cannot explain
$\Kr>1$ in the spinful case. On the other hand, the key difference
is the additional energy scale of the spin gap arising from
attractive backscattering in the spinful case. We will return to
this point in Sec.~\ref{sec:simhubb}.

\subsection{Charge susceptibility}\label{sec:chargesusceptibility}

\begin{figure}
  \includegraphics[width=0.5\textwidth]{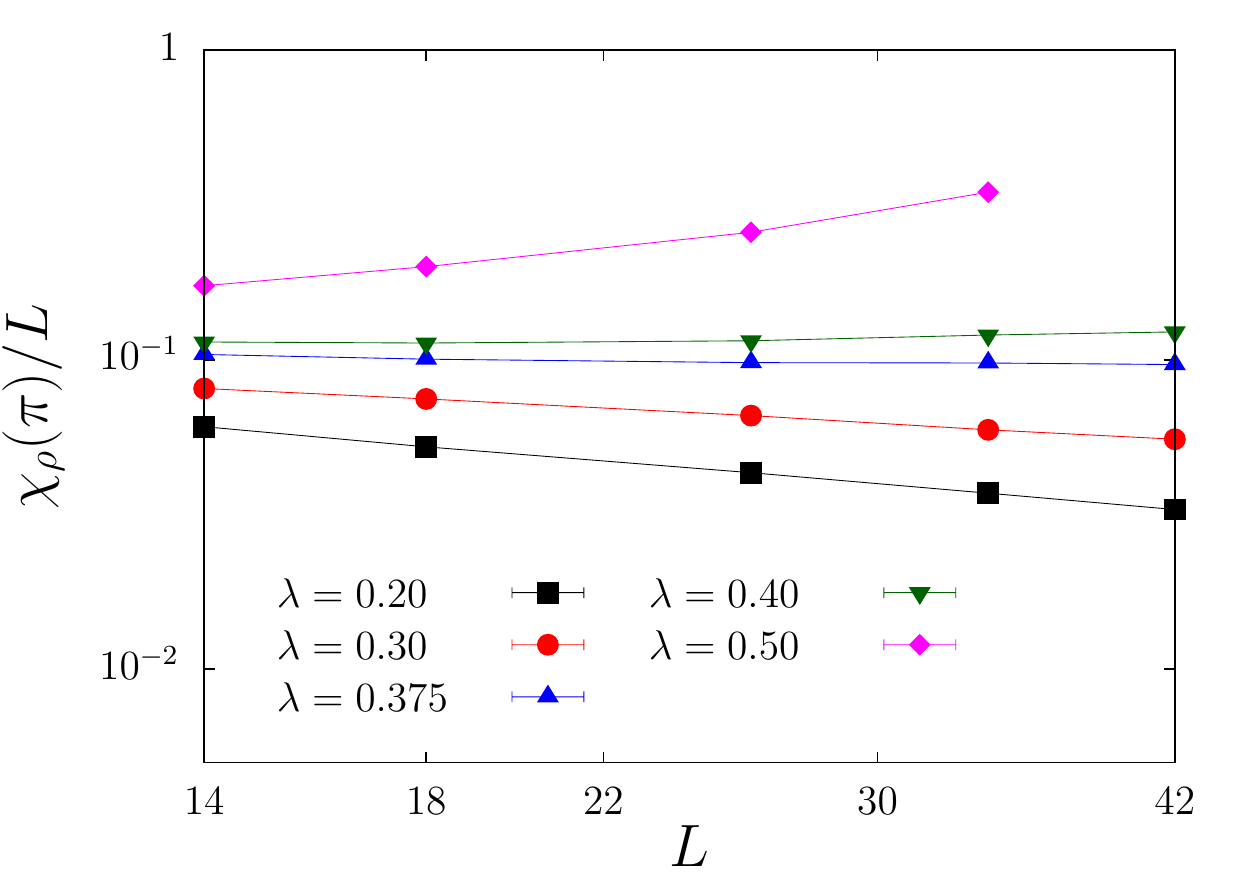}
  \caption{\label{fig:chiscalingspinless} (Color online) 
    Finite-size scaling of the charge susceptibility $\chi_\rho(\pi)$ [Eq.~(\ref{eq:chic})] for the spinless Holstein model.
    Here, $\om_0/t=0.1$, $\beta t=L$. Results obtained with the CT-INT method.
}
\end{figure}

While the onset of long-range order can be tracked using a finite-size
scaling of charge correlations, as shown in Fig.~\ref{fig:longrangeorder},
previous QMC studies of the phase diagram mainly relied on the charge
susceptibility 
\begin{equation}\label{eq:chic}
  \chi_\rho(Q) = \sum_{r=1}^L e^{i Q r} \int_0^\beta d\tau \las \on_r(\tau) \on_0(0) \ras
\end{equation}
at the ordering wavevector $Q=2\kF=\pi$. In principle, the
susceptibility has a more favorable scaling (a faster divergence in the
ordered phase) with system size
\cite{PhysRevLett.83.195,PhysRevB.67.245103}. However, because $\chi_\rho$
involves the charge correlation function, it is also affected by the spin gap.
We therefore discuss the expected and the observed behavior of the charge
susceptibility for Holstein models. 
To compare to existing work, we consider a phonon frequency $\om_0/t=0.1$ in
the spinless case, and $\om_0/t=0.5$ in the spinful case. 
Results for the $t$-$V$ and the
$U$-$V$ extended Hubbard model will be shown in Sec.~\ref{sec:simhubb}.

As shown in the appendix, assuming a power-law decay of $2\kF$
charge correlations of the form $(-1)^r r^{-\alpha}$, the susceptibility
scales as
\begin{equation}\label{eq:chiscaling}
  \chi_\rho(\pi)/L \sim C L^{1-\alpha}\,.
\end{equation}
Let us analyze the behavior of $\chi_\rho(\pi)/L$ for Luttinger
and Luther-Emery liquids, focusing on $K_\rho\leq 1$.

For a {\it spinless Luttinger liquid}, such as the spinless Holstein model
for $\lambda<\lambda_c$,  we have $\alpha=2K_\rho$ [see
Eq.~(\ref{eq:correl:LLspinless})]. At half-filling, umklapp scattering is
irrelevant for $K_\rho\geq1/2$, and relevant for $K_\rho<1/2$. 
Equation~(\ref{eq:chiscaling}) suggests the following behavior: for
$K_\rho=1$, $\chi_\rho(\pi)/L \sim \ln L/L$ and hence $\chi_\rho(\pi)/L\to 0$
for $L\to\infty$. Similarly, $\chi_\rho(\pi)/L\to 0$ for $1/2<K_\rho<1$. At
the critical value $K_\rho=1/2$, we have $\chi_\rho(\pi)/L\to C$. For
$K_\rho<1/2$, $\chi_\rho(\pi)/L$ diverges as a function of $L$. In
particular, $\chi_\rho(\pi)/L\sim L$ for $K_\rho=0$ (corresponding to long-range charge order). Hence, we should see a crossover in the
behavior of  $\chi_\rho(\pi)/L$ as a function of $K_\rho$ (or, equivalently,
the interaction strength), and can identify the critical point from the onset
of diverging behavior. The above considerations are borne out by the numerical results for the spinless Holstein
model~(\ref{eq:model-holsteinspinless}) shown in
Fig.~\ref{fig:chiscalingspinless}. Despite the limited system sizes, the  critical value
$\lambda_\text{c}\approx0.40$ agrees well with large-scale DMRG results \cite{Hohenadler06,Ej.Fe.09}.

For a {\it spinful Luttinger liquid} $\alpha=K_\rho+K_\sigma$
[Eq.~(\ref{eq:correl:LL})]. If $K_\sigma=1$ by symmetry, then
$\chi_\rho(\pi)/L\sim L^{-K_\rho}$. Therefore, we expect
$\chi_\rho(\pi)/L\to0$ for $0<K_\rho<1$ (the range where correlations decay
with a power-law), and a divergence of $\chi_\rho(\pi)/L$ in the ordered
state where the correlations approach a finite value at large distances.

\begin{figure}
  \includegraphics[width=0.5\textwidth]{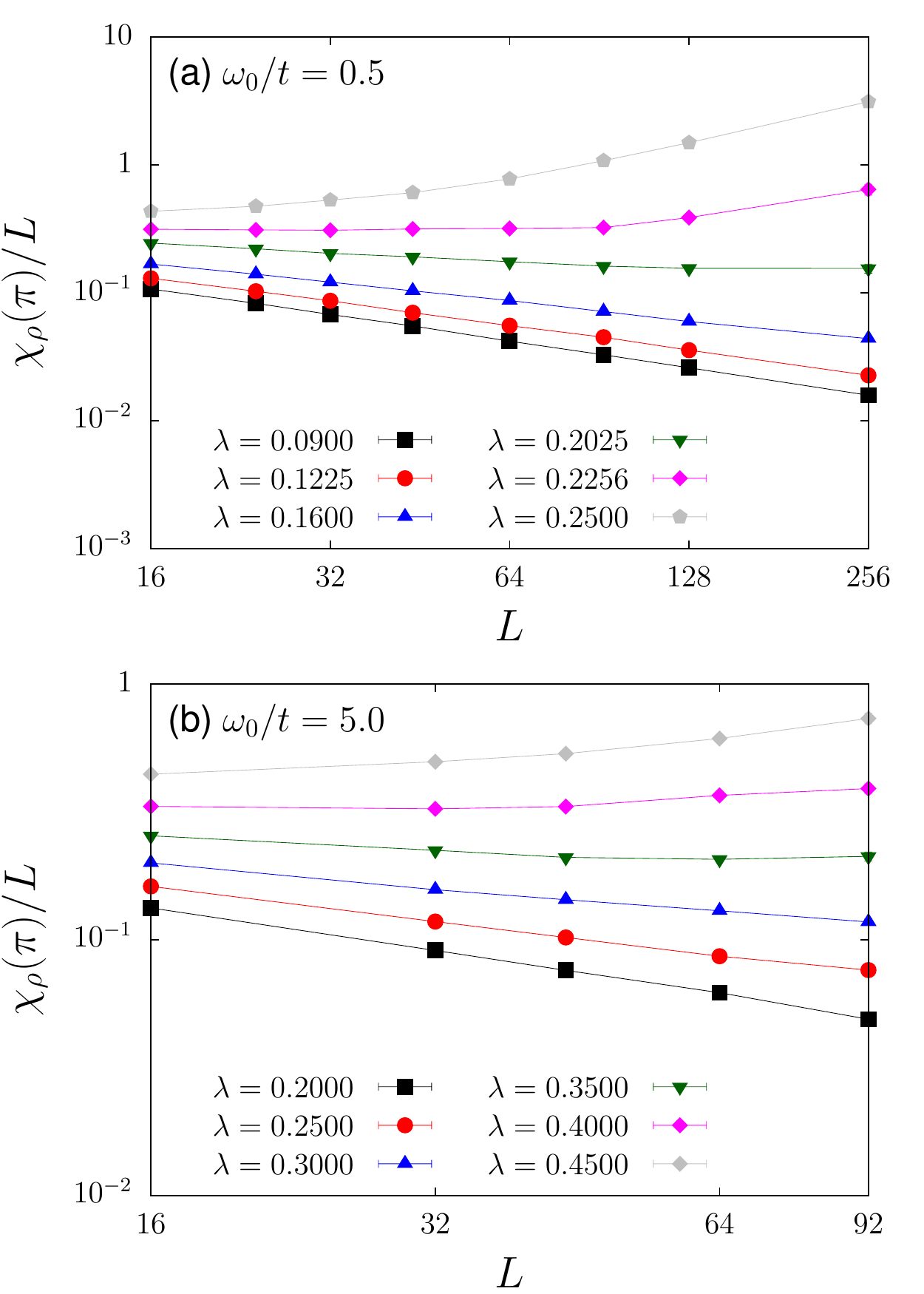}
  \caption{\label{fig:chiscalingspinful} (Color online) 
    Finite-size scaling of the charge susceptibility $\chi_\rho(\pi)$ for the
    spinful Holstein model for (a) $\omega_0/t=0.5$, and (b) $\omega_0/t=5$.
    Here, $\beta t=L$. Results obtained in the SSE representation.
}
\end{figure}

Finally, for a {\it Luther-Emery liquid}, as realized in the metallic phase
of the spinful Holstein model, $2\kF$ charge correlations decay with
$\alpha=K_\rho$ [Eq.~(\ref{eq:correl:LE}]. Consequently,
Eq.~(\ref{eq:chiscaling}) gives $\chi_\rho(\pi)/L \to C$ for $K_\rho=1$,
but $\chi_\rho(\pi)/L\sim L^{\epsilon}$ with $\epsilon=1-\alpha>0$ and hence 
$\chi_\rho(\pi)/L \to \infty$ for any $K_\rho<1$. Therefore, a metallic phase
with $K_\rho<1$ (dominant charge correlations, as observed for the Holstein
model) cannot be distinguished from a long-range-ordered Peierls phase from 
the qualitative behavior of $\chi_\rho(\pi)/L$ alone. This complication
appears to have been overlooked in Refs.~\onlinecite{ClHa05,hardikar:245103,Ho.As.Fe.12} where $\chi_\rho(\pi)$ was
used to track the Peierls transition in spinful models. The scaling
$\chi_\rho(\pi)\sim L^{2-K_\rho}$ for a Luther-Emery liquid was given in
Ref.~\onlinecite{Sa.Ba.Ca.04}. In principle, the Luther-Emery  and 
CDW phases may be distinguished by the different divergences of
$\chi_\rho(\pi)/L$. Given reliable estimates of $K_\rho$, the Luther-Emery
phase can be identified by plotting $\chi_\rho(\pi) L^{1-K_\rho}$, which
approaches a constant at large $L$ \cite{Sa.Ba.Ca.04}.

Let us compare these predictions to numerical data for the spinful Holstein model.
The crossover from Luttinger liquid to Luther-Emery liquid behavior in finite
systems corresponds to a change of the exponent $\alpha$ from $K_\rho+K_\sigma$ to $K_\rho$.
In the case of the Holstein model, the
charge correlation functions in Fig.~\ref{fig:powerlawspinful}(a) reveal an exponent $1<\alpha<2$ (the solid
line indicates $\alpha=2$, the dashed line $\alpha=1$) even though
the dominance of charge over pairing correlations implies $K_\rho<1$. This
suggests that on the length scales accessible in our results, $K_\sigma$ has
not scaled to zero, putting us between the two fixed points. 

The problems in resolving the spin gap and the correct long-distance behavior
of the correlation functions also affect the charge susceptibility. 
In contrast to the analytical predictions of a divergent  $\chi_\rho(\pi)/L$
for any $K_\rho<1$ in a Luther-Emery liquid, the data for $\chi_\rho(\pi)$ 
in Fig.~\ref{fig:chiscalingspinful}(a) strongly resemble the
spinless case shown in Fig.~\ref{fig:chiscalingspinless}. For small values of
$\lambda$, $\chi_\rho(\pi)/L\to 0$, whereas for sufficiently large
$\lambda$ it diverges. This behavior is consistent with previous work
\cite{ClHa05,hardikar:245103,Ho.As.Fe.12}, cf. Fig.~3(a) in
Ref.~\onlinecite{hardikar:245103}. 

Based on the incorrect assumption 
of $\chi_\rho(\pi)/L\to 0$ in the metallic phase, the phase boundary for the
Peierls transition was determined in Ref.~\onlinecite{hardikar:245103}. Similarly,
from Fig.~\ref{fig:chiscalingspinful}(a), we may (incorrectly) estimate
$\lambda_c\approx0.23$, in accordance with Ref.~\onlinecite{hardikar:245103}.
The DMRG estimate (from the opening of the two-particle gap) is
$\lambda_c\approx0.25$ \cite{0295-5075-84-5-57001}. For $\omega_0/t=1$, the transition was reported to occur at
$\lambda_c\approx0.25$ based on $\chi_\rho(\pi)/L$ \cite{hardikar:245103},
whereas a rough DMRG estimate (from the order parameter) is $\lambda_c\approx0.3$ \cite{JeZhWh99}. 
The spurious crossover in the behavior of
$\chi_\rho(\pi)/L$ is also apparent in Fig.~\ref{fig:chiscalingspinful}(b)
for $\omega_0/t=5$. Similar data were used before to 
estimate $\lambda\approx0.5$ \cite{hardikar:245103}. A comparison of QMC and
DMRG results can be made for the Holstein-Hubbard model with  $\omega_0/t=5$
and $U/t=1$. The QMC critical value is $\lambda_c\approx0.65$
\cite{hardikar:245103}, whereas DMRG estimates are $\lambda_c \approx
0.8$\,--\,1.0 \cite{0295-5075-84-5-57001} and $\lc\approx0.75$ \cite{1742-6596-200-1-012031}.

Our discussion reveals that the Peierls critical point for the spinful Holstein model
(or other models exhibiting a transition from a Luther-Emery to a CDW phase)
cannot be determined from a qualitative analysis (divergent or not) of
$\chi_\rho(\pi)/L$ because the latter diverges in both phases on sufficiently
large systems. Even a quantitative analysis of the divergent behavior will be
affected by the spin gap. Since the latter partially suppresses the
divergence of $\chi_\rho(\pi)/L$ for small $L$, critical values obtained from
the susceptibility are expected to be larger than the true values.
Given the importance of the Holstein-Hubbard model for our understanding of
electron-phonon physics, it is highly desirable to use alternative methods to
improve the accuracy of the phase diagram. 

\section{Fermionic models}\label{sec:simhubb}

In this section, we consider purely electronic models of spinful and spinless
fermions that exhibit a transition from a metallic phase to a
CDW insulator. This allows us to separate retardation effects
from those of backscattering (leading to a spin gap) and umklapp scattering (leading to a charge gap). 
The availability of exact analytical results provides a stringent test for
the numerical methods also used to study electron-phonon models. After
demonstrating that the spinless fermion problem is well accessible
numerically, leading to a very good agreement with analytical results, we will
show that the extended Hubbard model exhibits many of the issues encountered
in numerical simulations of the spinful Holstein model. Because of the better
scaling with system size, all results of this section were obtained in the
SSE representation.

\subsection{Spinless $t$-$V$ model}\label{sec:spinlessfermions}

The model of spinless fermions defined by Eq.~(\ref{eq:model-spinless})
captures the Luttinger liquid to CDW insulator transition also observed for
the spinless Holstein model. Importantly, it does not involve any retardation
effects, and can be solved exactly by the Bethe ansatz \cite{Bethe31}, thereby providing full
knowledge of the correlation functions and the phase diagram (see
Ref.~\onlinecite{PhysRevB.86.155156} and references therein).

\begin{figure}
  \includegraphics[width=0.5\textwidth]{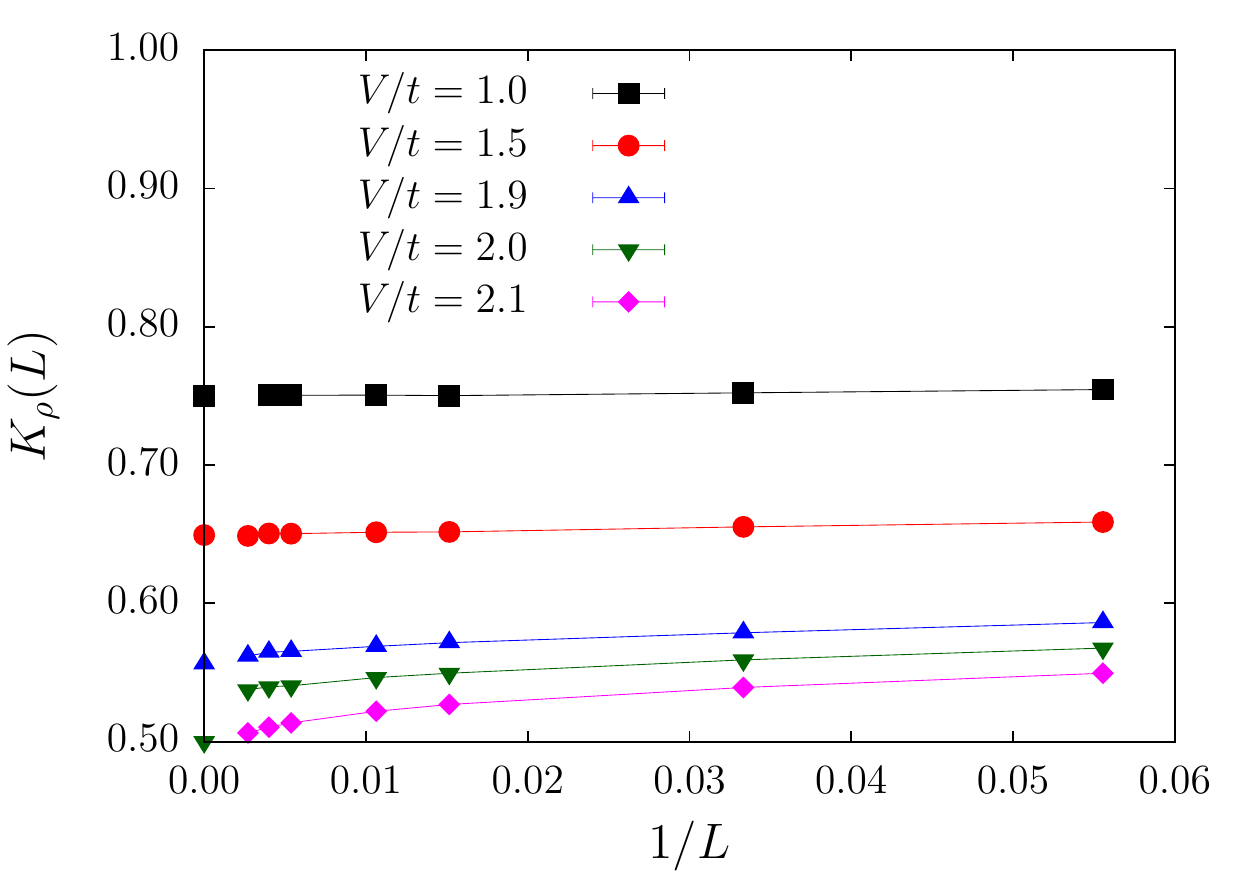}
  \caption{\label{fig:spinlesstV1} (Color online) 
    (a) $\Kr(L)$ [Eq.~(\ref{eq:krhoL})] for the spinless $t$-$V$ model. Here, $\beta t=2L$. The values for $L=\infty$
    were obtained from Eq.~(\ref{eq:Krhospinless}). Results obtained in the
    SSE representation.
}
\end{figure}

Figure~\ref{fig:spinlesstV1} shows $K_\rho(L)$ as a function of inverse
system size. In the metallic phase, which exists for $V/t<2$, the numerical
values extrapolate to the exact values \cite{PhysRevB.56.9766}
\begin{equation}\label{eq:Krhospinless}
  K_\rho = \frac{\pi}{2} \frac{1}{\arccos\left(-\frac{V}{2t}\right)}\,.
\end{equation}
Close to the transition, the
extrapolation becomes more difficult. In particular, $K_\rho$ should take on the
value $0.5$ exactly at the critical point $V/t=2$, and zero for
$V/t>2$. Similar results were obtained in a previous DMRG study \cite{0295-5075-70-4-492},
as well as for the spinless Holstein model \cite{Ej.Fe.09}.

\begin{figure}
  \includegraphics[width=0.5\textwidth]{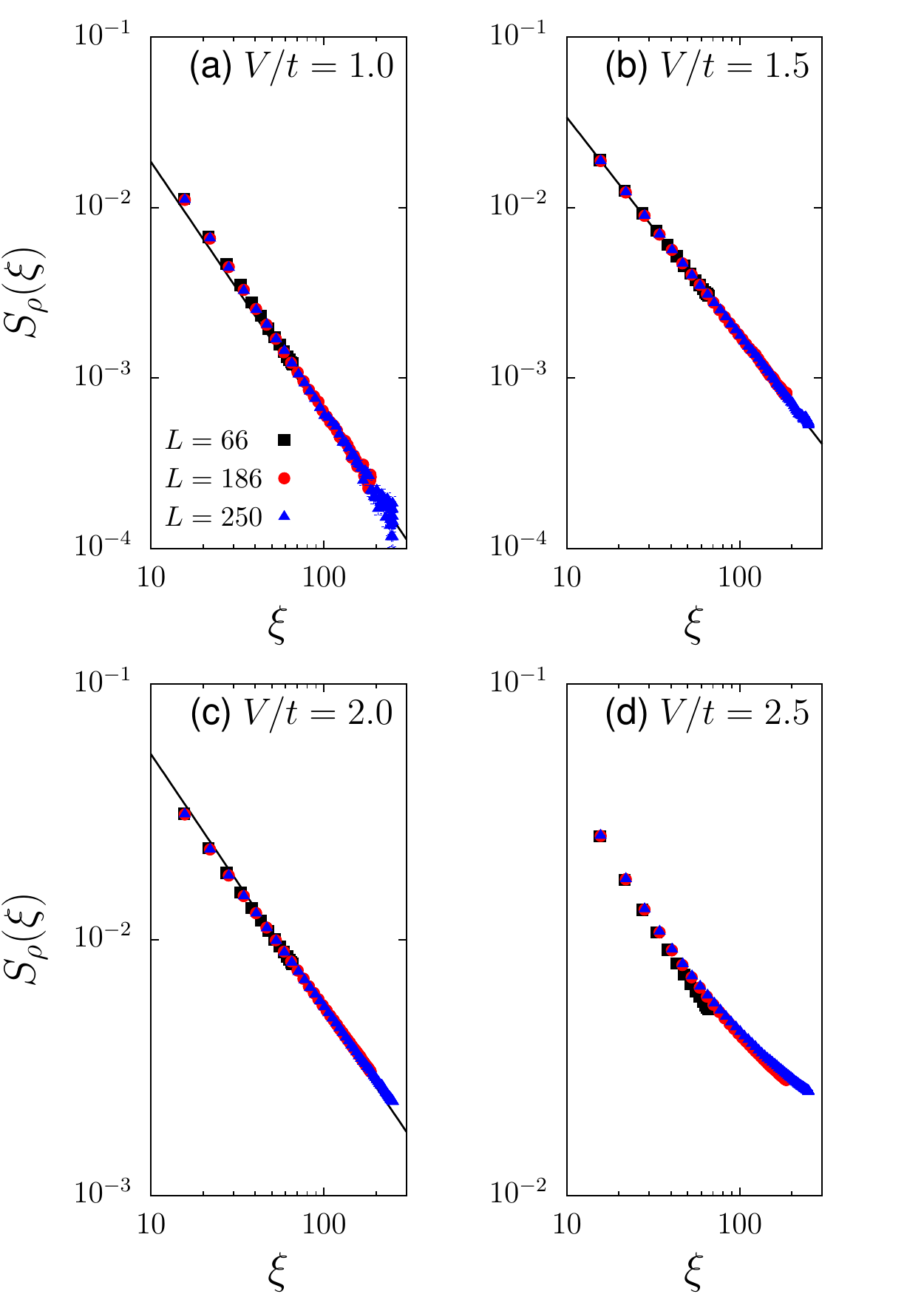}
  \caption{\label{fig:spinlesstV2} (Color online) 
    Charge correlation functions for the spinless $t$-$V$ model. 
    Solid lines correspond to fits to $a/x^{2\Kr}$ using results for $r>10$
    and $L=250$, with $\Kr$ from  Eq.~(\ref{eq:Krhospinless}).
    Here, $\beta  t=2L$. Results obtained in the SSE representation.
}
\end{figure}

The charge correlation function $S_\rho(\xi)$ is shown for different
values of $V/t$ in Fig.~\ref{fig:spinlesstV2}. In the metallic phase
($V/t<2$), we see a power-law decay for all system sizes considered.
Apart from the expected deviations at very small distances, the data
match the form $x^{-2K_\rho}$ [Eq.~(\ref{eq:correl:LLspinless})] with
$K_\rho$ taken from Eq.~(\ref{eq:Krhospinless}). Even exactly at the critical point $V/t=2$, the
numerical results do not indicate logarithmic corrections on the length scales
considered. For $V/t=2.5$, we see long-range order. Similar to the
Holstein models, we have a collapse of results for different $L$ in the
metallic phase [Figs.~\ref{fig:spinlesstV2}(a)--(c)], but not in the
insulating phase [Fig.~\ref{fig:spinlesstV2}(d)]. 

\begin{figure}
 \includegraphics[width=0.5\textwidth]{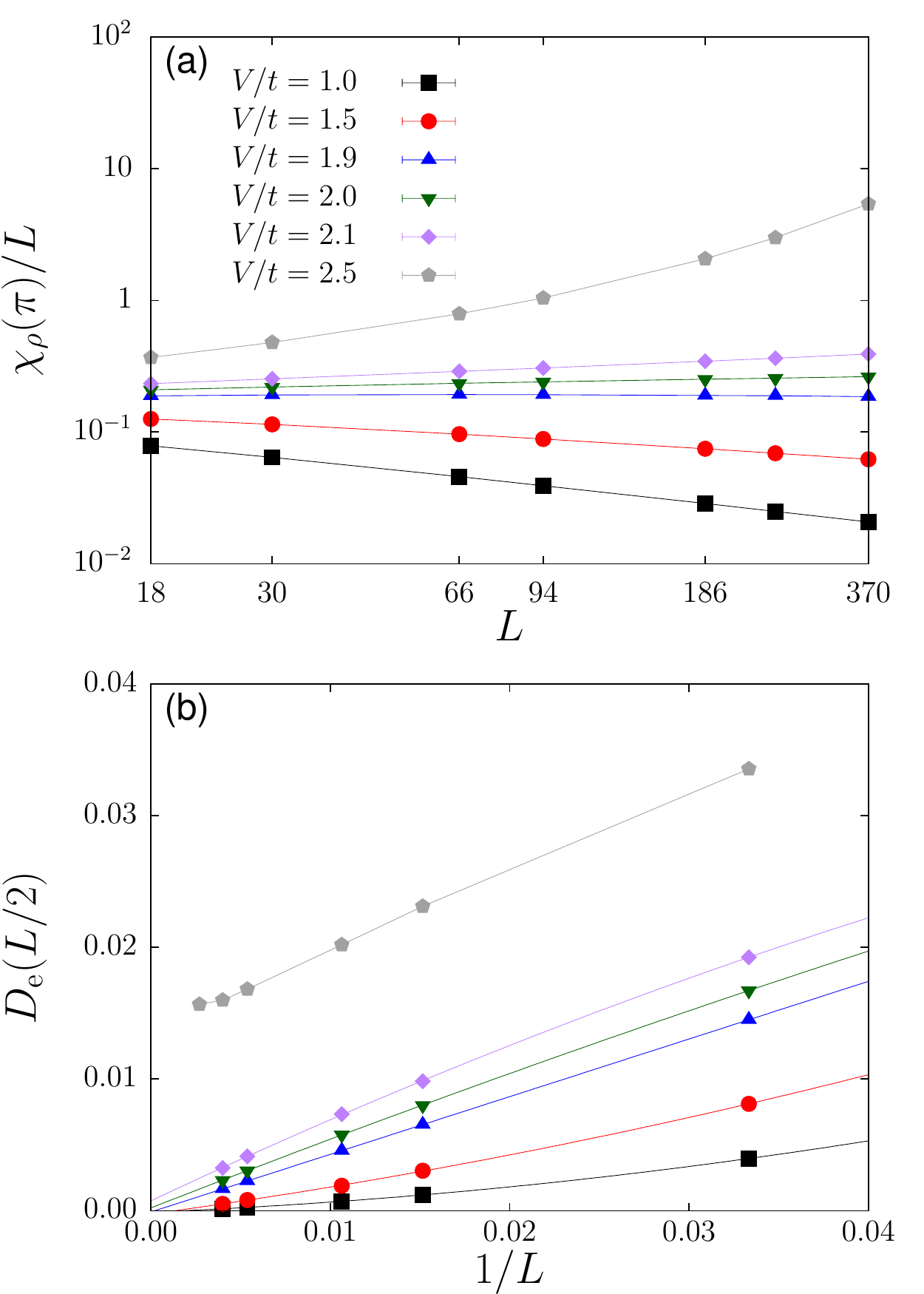}
  \caption{\label{fig:spinlessV3} (Color online) 
   Finite-size scaling of (a) the charge susceptibility and (b) the
   order parameter for the spinless $t$-$V$ model. Here, $\beta t=2L$. Results obtained in the
    SSE representation.
  }
\end{figure}

The charge susceptibility $\chi_\rho(\pi)$ is shown in Fig.~\ref{fig:spinlessV3}(a). 
Since $S_\rho(\xi)$ behaves as expected from the
bosonization, the numerical results for $\chi_\rho(\pi)/L$ are consistent
with the discussion in Sec.~\ref{sec:chargesusceptibility}. For $V/t<2$,
$\chi_\rho(\pi)/L$ goes to zero with increasing system size, whereas it 
diverges for $V/t>2$. The susceptibility can therefore be used to
determine the critical value of the transition even on moderately large
systems. The same conclusion holds for the spinless Holstein model.

Finally, we turn to the charge order parameter as an alternative way to
track the phase transition. To this end, we consider the charge correlations
at the largest distance $L/2$ for different system sizes $L$, shown in 
Fig.~\ref{fig:spinlessV3}(b). In the metallic phase, this quantity is expected
to be zero in the thermodynamic limit, and nonzero in
the charge-ordered insulating phase. 
The data in Fig.~\ref{fig:spinlessV3}(b) show that the scaling of the
charge correlations allows to quite accurately determine the critical value
of the phase transition. The same is true for the extended Hubbard model (not
shown), and we therefore expect this order parameter to be a reliable tool to
determine the transition for electron-phonon models.

\subsection{Attractive Hubbard model}\label{sec:attractivehubbard}

In Sec.~\ref{sec:holstein} we identified the existence of a (small) spin gap
in the spinful Holstein model as a significant complication in finite-size
studies. Because the retarded interaction and the nonintegrability of the
Holstein model make further progress difficult, we turn to the exactly
solvable attractive Hubbard model [Eq.~(\ref{eq:model-hubbard}) with $V=0$].

At half-filling, the attractive Hubbard model is related by a canonical transformation to the repulsive Hubbard
model. From bosonization/RG studies \cite{Emery79}, it is known that its ground state
is metallic for any $U<0$ and has a gap for spin excitations. It therefore provides a
numerically accessible realization of the Luther-Emery fixed point. The
absence of phonons permits significantly larger system
sizes to be simulated, and eliminates any uncertainties from the
bosonization/RG treatment related to a phonon energy scale.

The exact relation to the repulsive Hubbard model has important
consequences. The value of $K_\rho$ is fixed to 1 for any
$U<0$ because $K_\rho$ in the attractive model corresponds to
$K_\sigma$ in the repulsive model [where $K_\sigma=1$ because of SU(2) spin symmetry]. Additionally, $K_\sigma=0$ as a
result of the spin gap. A value $K_\rho=1$ implies that
charge and pairing correlation functions remain degenerate for any $U$
[cf. Eq.~(\ref{eq:correl:LE})], and their $2\kF$ components are expected to decay
as $x^{-1} \sqrt{\text{log}(x)}$ \cite{Schulz90,PhysRevB.45.4027}. 

\begin{figure}
  \includegraphics[width=0.5\textwidth]{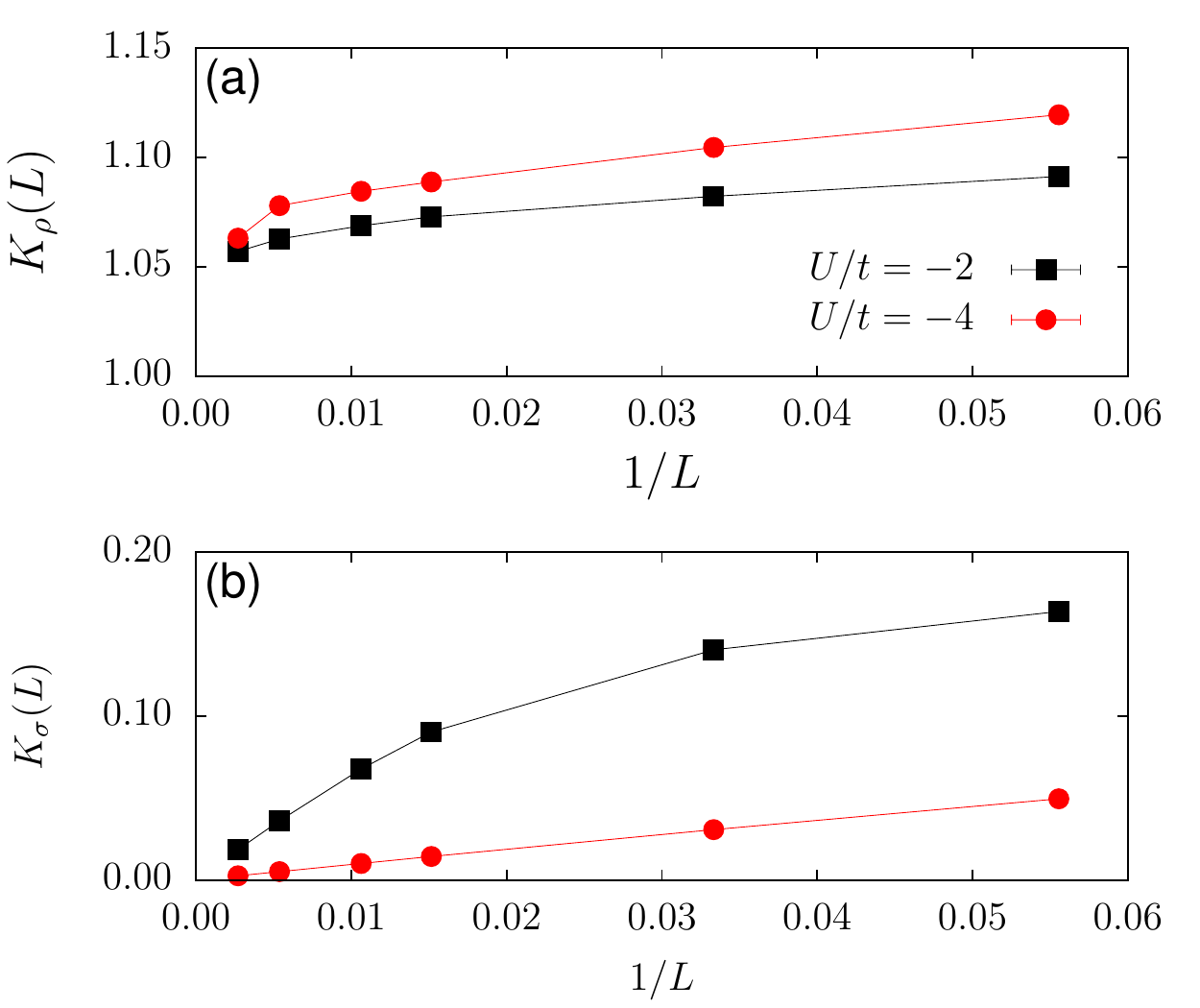}    
  \caption{\label{fig:attractivehubbard1} (Color online) 
    (a) $\Kr(L)$ and (b) $\Ks(L)=2\pi
    S_\sigma(q_1)/q_1$ for the attractive Hubbard model. Here, $\beta t=2L$. Results obtained in the
    SSE representation.
}
\end{figure}

\begin{figure}
  \includegraphics[width=0.5\textwidth]{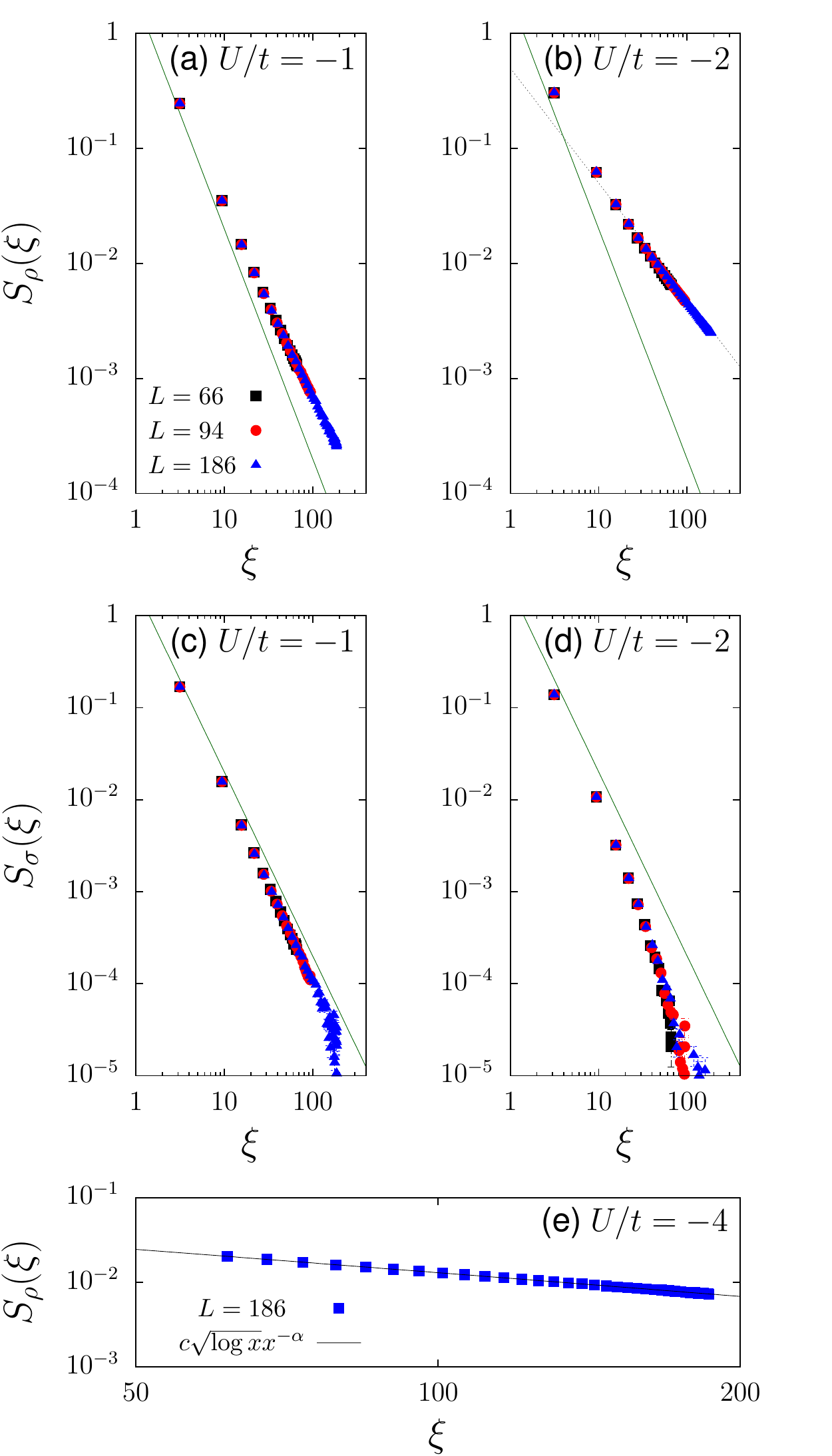}
  \caption{\label{fig:attractivehubbard2} (Color online) 
    Charge and spin correlation functions for the attractive Hubbard
    model. In (a)\,--\,(d), solid (dashed) lines illustrate $c/x^2$
    ($c/x$). Here, $\beta  t=2L$. Results obtained in the  SSE representation.
    }
\end{figure}

Figure~\ref{fig:attractivehubbard1}(a) shows $\Kr(L)$ from simulations in the
SSE representation on up to {$L=370$} sites. In agreement with previous
findings \cite{hardikar:245103}, $\Kr(L)$ is significantly larger than 1, and
for a given $L$ increases with increasing $U$, reminiscent of the increase of
$K_\rho(L)$ with increasing $\lambda$ in
Fig.~\ref{fig:krhofromdensitylambda}(a). The convergence with system size is
very slow, making it challenging to obtain $K_\rho=1$ from an unbiased extrapolation.
The relation between $\Kr$ of the attractive and
$\Ks$ of the repulsive Hubbard model is reflected in the observation that
$\Ks(L)>1$, with a very slow convergence to the value $\Ks=1$ implied by
symmetry (data not shown).  A similarly slow convergence has also been observed for the
repulsive Hubbard model with open boundaries, where $K_\sigma(L)$ was
determined from fits to the local density of states \cite{0295-5075-101-5-56006}.
In the attractive case considered here,
Fig.~\ref{fig:attractivehubbard1}(b) shows that $\Ks(L)<1$, and that the
numerical data are consistent with $\Ks=0$. Similar results have been obtained before for the Holstein
model \cite{PhysRevB87.075149}. Values $\Ks(L)<1$ in numerical simulations
are a reliable indicator for the existence of a spin gap
\cite{PhysRevB.59.4665}. 

Given the slow convergence of $K_\rho(L)$ for the attractive Hubbard model, 
the reliability of estimates of $K_\rho$ for the spinful Holstein model has to be questioned.
As mentioned above, the slow convergence may be attributed to unknown
logarithmic corrections \cite{0305-4470-22-5-015,PhysRevB.39.4620,Lukyanov1998533}. 
While such corrections arise in the repulsive
Hubbard model from the marginally irrelevant backscattering term, the umklapp
term is marginally relevant (since $K_\rho=1$ by symmetry) in the attractive model.
Logarithmic corrections are typically absent in spinless models, in
accordance with our findings. The deviations of $K_\rho(L)$ from the expected value
$K_\rho=1$ for a given $L$ are more pronounced for stronger interactions
(larger spin gaps), similar to
Ref.~\onlinecite{0295-5075-101-5-56006}. Finally, in a
Luther-Emery phase, the factor $C_\rho$ in the first term in $S_\rho$ in
Eq.~(\ref{eq:correl:LE}) may not be identical to $K_\rho$, and logarithmic
corrections may arise.

\begin{figure}
  \includegraphics[width=0.5\textwidth]{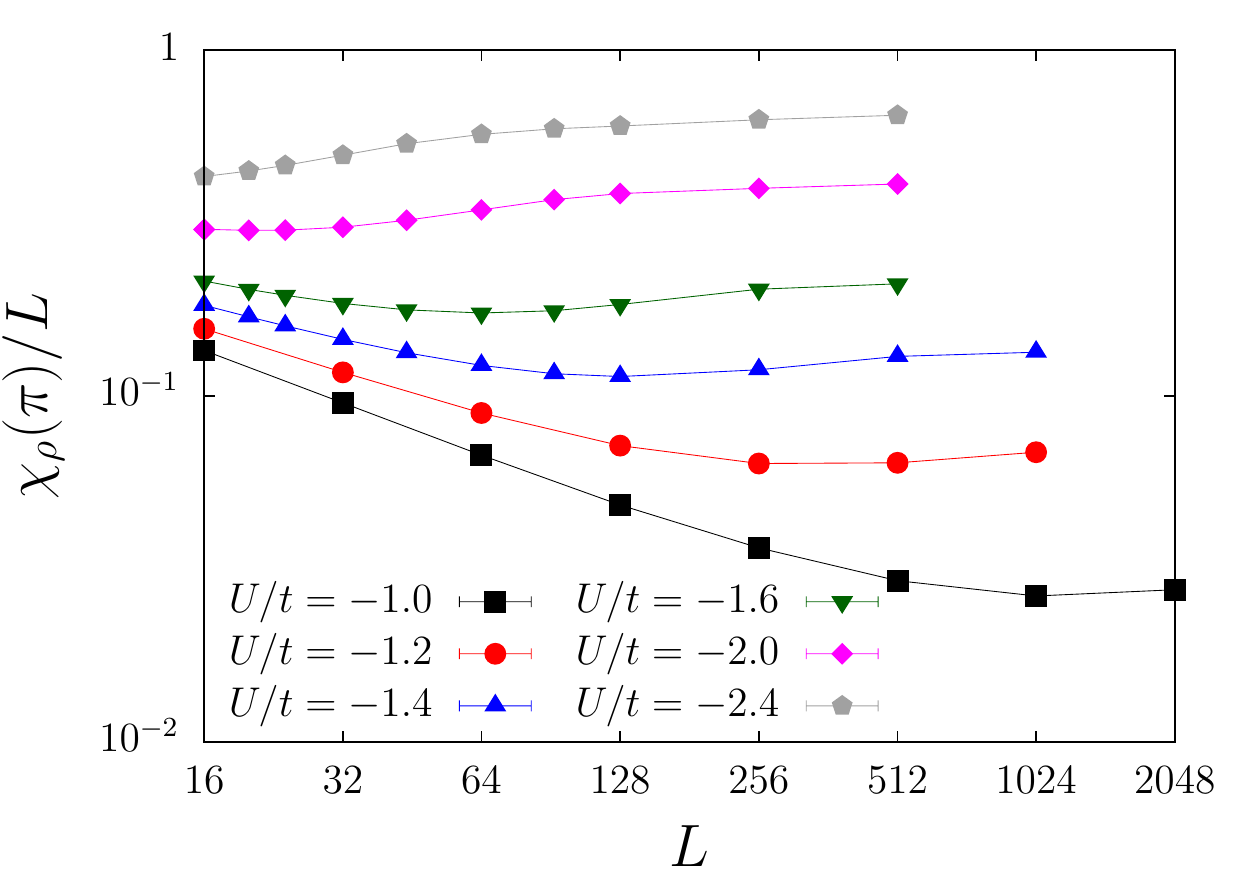}
  \caption{\label{fig:attractivehubbard3} (Color online) 
   Finite-size scaling of the charge susceptibility for the attractive Hubbard model.
   Here, $\beta t=2L$. Results obtained in the SSE representation.
  }
\end{figure}

Given a certain spin gap as a result of attractive backscattering, we 
expect to see physics reminiscent of the Tomonaga-Luttinger fixed
point for distances small compared to the inverse of the spin gap, and
Luther-Emery behavior at large distances. The significantly larger
system sizes accessible for the attractive Hubbard model permit us to illustrate this
fixed-point crossover by comparing the numerical correlation functions to the Luttinger
liquid and Luther-Emery results given in Eqs.~(\ref{eq:correl:LL})
and~(\ref{eq:correl:LE}), respectively.

For $U/t=-1$, the charge correlations 
shown in Fig.~\ref{fig:attractivehubbard2}(a) decay with 
an exponent close to $2$ at small distances, as expected from
Eq.~(\ref{eq:correl:LL}) for $K_\sigma=1$, $K_\rho=1$. With increasing
distance $\xi$, we observe a crossover of the exponent caused by
$K_\sigma\to0$. For $U/t=-2$, the spin gap is significantly larger, and the
crossover from $x^{-2}$ to approximately $x^{-1}$ is visible in Fig.~\ref{fig:attractivehubbard2}(b). Regardless of the
crossover, we observe a collapse of data for different system sizes over the
whole range of $\xi$.  The spin correlations for $U/t=-1$, shown in
Fig.~\ref{fig:attractivehubbard2}(c), decay almost as in a gapless system
(\ie, as $x^{-2}$), whereas for $U/t=-2$ the results are compatible with an
exponential decay [Fig.~\ref{fig:attractivehubbard2}(d)]. Hence, on length
scales smaller than the inverse spin gap, charge correlations decay faster
than in the thermodynamic limit, whereas spin correlations decay slower.
While such a crossover is also expected for the Holstein model, the
range of numerically accessible system sizes is insufficient to observe it
numerically. 

Figure~\ref{fig:attractivehubbard2}(e) shows the charge correlations for
$U/t=-4$, where the spin gap is large enough to be resolved even on
intermediate system sizes. Also shown is a  fit to the prediction
$c\sqrt{\log x} x^{-\alpha}$  from the bosonization (solid line) using data for
$L=186$ and $r\geq20$. The fit is in satisfactory agreement with the data;
we find an exponent $\alpha\approx1.03$ compatible with
Fig.~\ref{fig:attractivehubbard1} and indicative of possible 
logarithmic corrections
\cite{0305-4470-22-5-015,PhysRevB.39.4620,Lukyanov1998533,0295-5075-101-5-56006}
in addition to the crossover due to the spin gap.

The crossover in the correlation functions also affects the scaling of the
charge susceptibility. If $S_\rho(x)$ decays faster than $x^{-1}$,  $\chi_\rho(\pi)/L\to 0$ for
$L\to\infty$. This is well visible in Fig.~\ref{fig:attractivehubbard3} for
small $U/t$. For example, for $U/t=-1$,  $\chi_\rho(\pi)/L$ decreases with
increasing $L$ up to $L=1024$, well beyond the system sizes accessible for
electron-phonon models. If the decay is exactly $x^{-1}$, $\chi_\rho(\pi)/L$
should approach a constant at large $L$. However, the 
logarithmic correction $\sqrt{\log x}$ for the attractive Hubbard model gives rise
to a slow divergence of $\chi_\rho(\pi)/L$ with system size, as visible in
Fig.~\ref{fig:attractivehubbard3} at large $L$. The different behavior of
$\chi_\rho(\pi)/L\to 0$ for different $U/t$ in
Fig.~\ref{fig:attractivehubbard3} may be mistaken as evidence for a
transition to a CDW-ordered phase, even though the model remains
metallic. Similar deviations from the expected behavior of
$\chi_\rho(\pi)/L$ on finite systems can be observed for the spinful Holstein model
(see Fig.~\ref{fig:chiscalingspinful}).

\begin{figure}
  \includegraphics[width=0.5\textwidth]{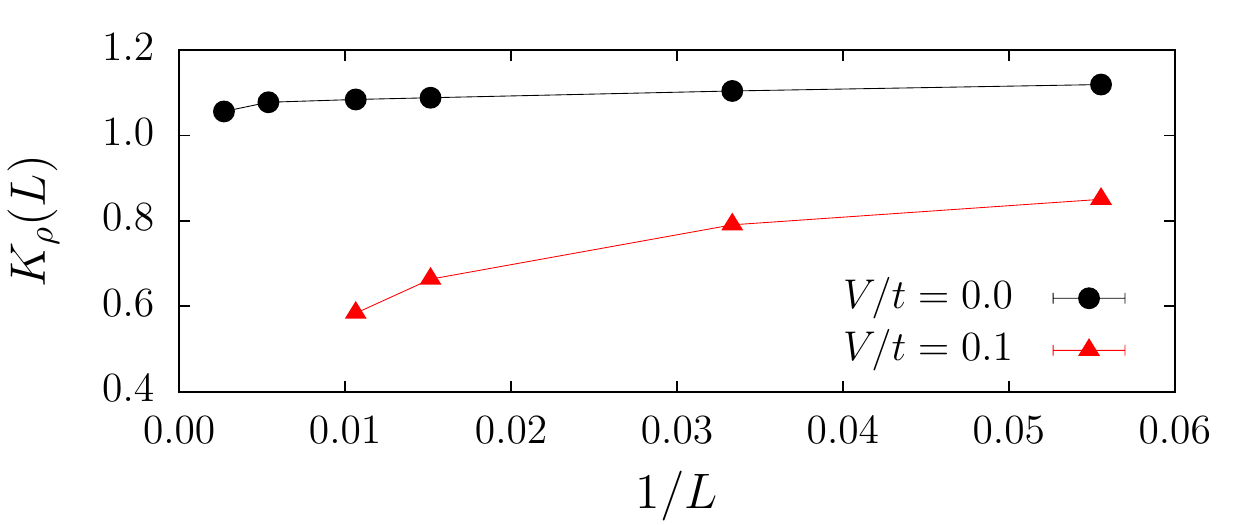}
  \caption{\label{fig:UV1} (Color online) 
    $\Kr(L)$ for the $U$-$V$ model. Here, $\beta t=2L$, $U/t=-4$. Results obtained in the
    SSE representation.
}
\end{figure}

\subsection{Extended Hubbard model}\label{sec:extendedhubbard}

Whereas the attractive Hubbard model is always a Luther-Emery metal, we can study the transition to a charge-ordered insulating phase by
adding a repulsive nearest-neighbor interaction $V\sum_i \on_i \on_{i+1}$, see
Eq.~(\ref{eq:model-hubbard}). This extended Hubbard model captures the
existence of a spin gap in the metallic phase and $2\kF$ CDW order
in the insulating phase, similar to the Holstein model, while excluding
retardation effects. The aforementioned strong-coupling approximation of
Ref.~\onlinecite{Hirsch83a} (cf. Sec.~\ref{sec:overview}) suggests that the low-energy physics of the
Holstein model is described by a model of bosonic pairs that hop and interact repulsively.
Whereas the critical value for the metal-insulator transition is $V=0$ in the
electronic model, the Holstein model supports an extended metallic region. 

Figure~\ref{fig:UV1} shows $K_\rho(L)$ as a function of $1/L$. The
results for $V=0$ are identical to those in
Fig.~\ref{fig:attractivehubbard1}(a), and extrapolate to $K_\rho=1$. In
contrast, for $V/t=0.1$, corresponding to the insulating phase with long-range
charge order, the results for $K_\rho(L)$ are compatible with $K_\rho=0$.

\begin{figure}
  \includegraphics[width=0.5\textwidth]{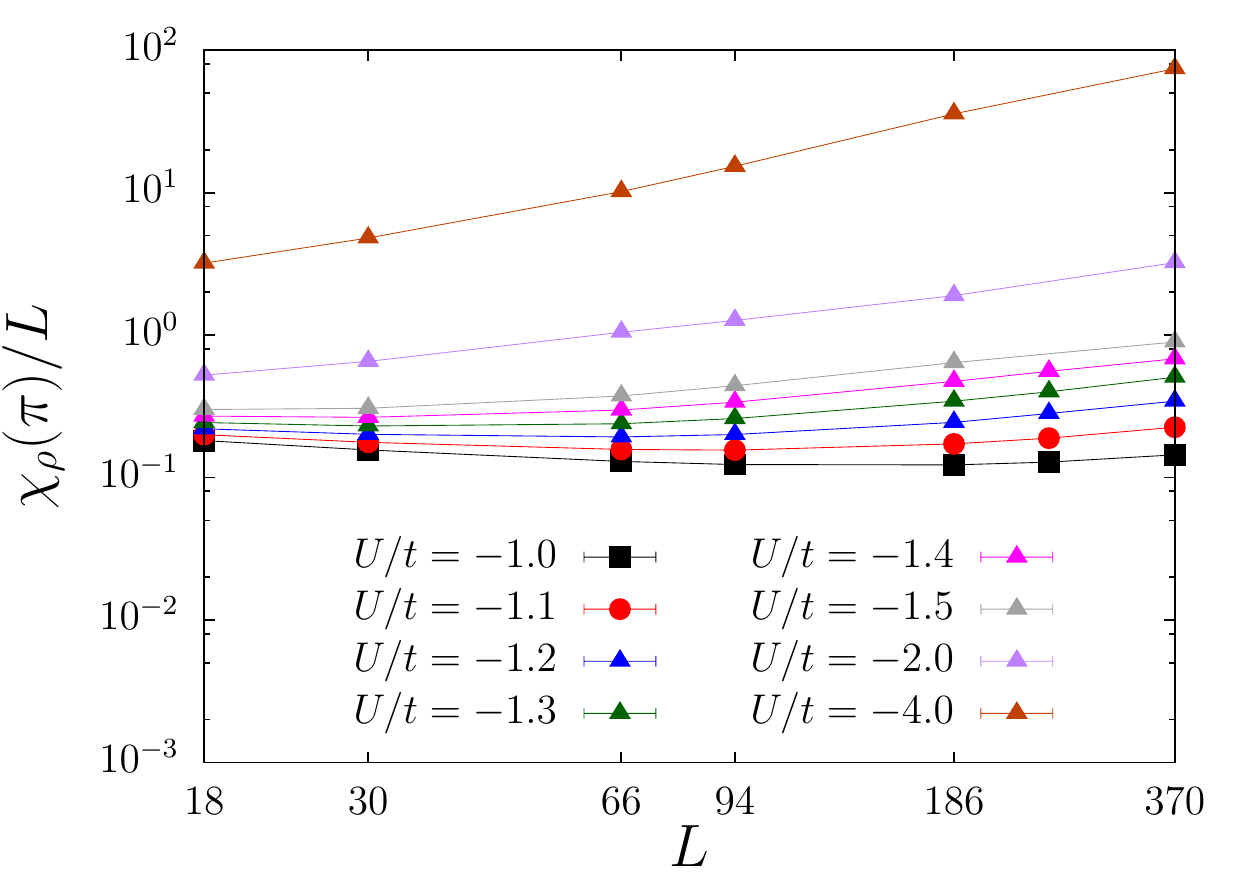}
  \caption{\label{fig:UV2} (Color online) 
   Finite-size scaling of the charge susceptibility for the $U$-$V$ model.
   Here, $\beta t=2L$, $V/t=0.1$.  Results obtained in the
    SSE representation.
  }
\end{figure}

In Fig.~\ref{fig:UV2}, we show results for $\chi_\rho(\pi)/L$ in the
insulating CDW phase ($V/t=0.1$).  For $U/t=-1$, the spin gap 
is not resolved in the Luther-Emery phase for $V=0$ (see
Fig.~\ref{fig:attractivehubbard3}), which leads to a
nonmonotonic finite-size scaling in the CDW phase at $V/t=0.1$. For small $L$,
$\chi_\rho(\pi)/L$ decreases,
whereas for large $L$ the expected  divergence becomes visible.
Such nonmonotonic behavior can also be observed for the Holstein
model [see Fig.~\ref{fig:chiscalingspinful}(b) and
Ref.~\onlinecite{hardikar:245103}]. The minimum at intermediate $L$
persists for slightly larger $|U/t|$ (but is shifted to smaller $L$ because
the spin gap increases with increasing $|U/t|$), whereas for large $|U/t|$
the gap is large enough to be resolved even for small $L$. For these
parameters (\eg, $U/t=-4$), $\chi_\rho(\pi)/L$ is monotonic and exhibits a
clear divergence. Because $\chi_\rho(\pi)/L$ even diverges
(logarithmically) in the metallic phase for $V/t=0$ (see
Fig.~\ref{fig:attractivehubbard3}), its
qualitative behavior at large $L$ does not permit to distinguish the metallic
from the insulating phase.

\section{Discussion}\label{sec:discussion}

\subsection{Metallic phase}

Numerical results for the  correlation functions reveal that in the spinful
Holstein model at half-filling, charge correlations dominate over pairing
correlations down to very small values of the electron-phonon
coupling \cite{tam:161103,PhysRevB87.075149}. At the same time, spin
correlations are suppressed relative to charge
correlations \cite{PhysRevB87.075149}. For a 1D metallic system, this necessarily
implies a gap for spin excitations \cite{Voit98,PhysRevB87.075149},
consistent with the observation of $\Ks<1$ \cite{PhysRevB87.075149}.
The gap arises from the pairing of electrons into spin singlets (bipolarons)
or, in the language of bosonization, from attractive backscattering which 
is a relevant perturbation of the Tomonaga-Luttinger fixed point.

The spin gap makes the Luther-Emery liquid the relevant fixed
point for a low-energy description of the metallic phase, and significantly
complicates the analysis of numerical data because a crossover
from the Luttinger to the Luther-Emery fixed point takes place as a
function of distance. For distances smaller than the inverse spin gap,
correlation functions resemble those of a gapless Luttinger liquid, whereas
for large distances the corresponding Luther-Emery results are
approached. Because the spin gap is small in the metallic phase for typical
phonon frequencies, it appears highly nontrivial to reach the correct low-energy
fixed point numerically. Importantly, the exponents for charge and pairing
correlations are expected to change in the same way when the spin gap is
resolved, leaving the conclusion that the metallic phase is a repulsive
Luther-Emery liquid unchanged. The fixed-point crossover and the complications in
determining the Luttinger parameters and the phase boundary are absent in the
spinless Holstein and $t$-$V$ models.

Assuming that the spin gap in the Holstein model is roughly determined by the
corresponding attractive Hubbard interaction $U=-4\lambda t$, the critical values
$\lambda_\text{c}\approx0.25$ (for $\om_0/t=0.5$) and
$\lambda_\text{c}\approx0.5$ (for $\om_0/t=5$) translate into $U/t=-1$ and $-2$,
respectively. Hence, the results for the attractive Hubbard model in
Fig.~\ref{fig:attractivehubbard2} should be representative of the largest
spin gaps in the metallic phase of the Holstein model for typical phonon
frequencies, and suggest that the 
Luther-Emery regime will be hard to reach numerically (\eg, for $U/t=-1$, 
$L=370$ sites are not sufficient). If system sizes are too small,
correlation functions will not display the expected behavior.

The numerical finding of $\Kr>1$ in the metallic phase has motivated much of
the recent work on the Holstein model. Initial claims of dominant pairing
correlations turned out to be incorrect, and the system instead exhibits
dominant charge correlations suggestive of $\Kr<1$. We have shown that
the inconsistent values of $\Kr$ cannot be explained by 
retardation effects ($\Kr$ deviates even more from 1 for
higher phonon frequencies). Instead, because very similar problems arise in
simulations of the attractive Hubbard model, we attribute the complications
to backscattering and the spin gap.

\subsection{Low-energy theory}

While theories of metal-insulator transitions in Luther-Emery liquids
are less controlled than for Luttinger liquids \cite{PhysRevB.45.4027}, 
a consistent picture of the spinful Holstein model is in terms
of interacting bipolarons that order into a CDW state at
$\lambda_c$. Such a description  was previously suggested
for the extended Hubbard model \cite{Emery79}, and also emerges from a
strong-coupling approximation of the Holstein model \cite{Hirsch83a}. For simplicity,
we focus the discussion on hard-core bosons, although bipolarons in the
Holstein model can also be extended.

Adopting the bosonic picture, the bosonization and RG methods predict
metallic behavior (reflected in a vanishing two-particle electronic charge
gap) for $K>1/2$ (we drop the index $\rho$ when referring to
the bosonic picture). Similar to spinless fermions, a metallic phase with
dominant charge correlations and subdominant pairing correlations exists for
$1/2<K<1$. At $K=1/2$, a Mott transition to an insulating CDW state
takes place. Clearly, the bosonic picture provides a strong connection to the
spinless Holstein model, where the Luttinger liquid to Peierls insulator
transition may be understood in terms of a Mott transition of spinless fermions.

The value of $K_\rho$ at the Peierls transition in the spinful Holstein model
is not known from theory (the assumption of
$K_\rho=1$  in Ref.~\onlinecite{1742-6596-200-1-012031} seems unjustified because there is no
cancellation of interactions at $\lambda_c$). The transition from a
Luther-Emery to a CDW phase was studied for the $U$-$V$ extended Hubbard
model, where the symmetries for $V=0$ imply that the critical value is
$K_\rho=1$. However, the bosonic low-energy theory of the Holstein model
contains an interaction between pairs of electrons, and we therefore expect a
different scaling dimension and critical $K$. Additionally, the role
of quantum lattice fluctuations needs to be addressed. Indeed, recent functional RG
results for the Holstein model \cite{Barkim2015} confirm the existence of a
finite metallic region. 

For numerical investigations of the Holstein model, it is important to
recognize that the system size has to be sufficiently large to resolve the
spin gap in order to allow for a meaningful comparison with the bosonic
picture. The electronic Luttinger parameter $K_\rho$ may in general not have
a simple relation to the $K$ of the bosonic picture. Of particular interest
is a numerical calculation of bosonic pairing and density correlation
functions. The latter would permit to test the bosonic picture and the
corresponding bosonization results (including the transition at $K=1/2$)
quantitatively.

\section{Conclusions}\label{sec:conclusions}

We carried out extensive quantum Monte Carlo simulations to resolve conflicts
regarding the existence of a metallic phase and the value to the Luttinger
parameter $K_\rho$ in the one-dimensional spinful Holstein model at half-filling.
In addition, we considered the spinless Holstein model and minimal fermionic
models that capture the charge-density-wave transition. 

First, we showed that the results of
Ref.~\onlinecite{Hirsch83a} are partly incorrect (likely due to autocorrelations),
and should hence not be regarded as evidence for the absence of a
metallic phase in the Holstein model, in accordance with recent functional
renormalization group results that predict a nonzero critical electron-phonon
coupling \cite{Barkim2015}.

Our results for the real-space correlation
functions in the metallic phase are consistent with Luttinger liquid physics
for the spinless Holstein model, and Luther-Emery physics for the spinful
Holstein model. In the latter case, spin correlations are suppressed with
respect to charge correlations due to the existence of a spin gap, although
the gap is typically not fully resolved in numerical simulations.
In both models, charge correlations dominate over pairing correlations, which
suggests predominantly repulsive interactions.

Given a metallic phase at weak coupling, we investigated in detail how the
spin gap manifests itself in finite-size simulations. By comparing results
for the spinful and the spinless Holstein model, as well as for fermionic
models, we revealed that complications in accurately determining $K_\rho$
appear to be generic for spin-gapped phases, but independent of retardation effects.
As a function of distance,
correlation functions reveal a crossover from Luttinger liquid behavior at
short distances to Luther-Emery behavior at long distances. A nonzero spin
gap also affects the possibility of using the charge susceptibility to track the
Peierls transition, which may explain existing deviations between phase
boundaries in the literature.  Our results and analysis of previous work
suggest that even with state-of-the-art numerical methods the paradigmatic
spinful Holstein and Holstein-Hubbard models remain challenging. 

Finally, our work motivates further numerical and analytical
investigations. On the numerical side, it would be useful to find alternative
ways to determine the critical value of the Peierls transition, and  to
measure bosonic correlation functions to verify the low-energy description in
terms of bipolarons. On the theoretical side, an improved understanding of
the origin of logarithmic corrections to $K_\rho$, and of the Mott transition
in a generic Luther-Emery liquid would be desirable. 

{\begin{acknowledgments}%
We acknowledge computer time at the J\"ulich Supercomputing Centre,
financial support from the DFG via FOR 1807, and
valuable discussions with C. Bourbonnais, S. Eggert,
S. Ejima, F. Essler, H. Fehske, T. Giamarchi, V. Meden, A. Sandvik, and D. Schuricht.
\end{acknowledgments}}

\appendix

\section{Scaling of the charge susceptibility}\label{app:chic}

For $K_\rho<1$, the dominant contribution to the equal-time charge
correlations in a Luttinger or Luther-Emery liquid has $Q=2\kF$, with $Q=\pi$
for half-filling. Writing the equal-time $2\kF$ charge correlations
in the form $(-1)^r/r^\alpha$, and exploiting conformal invariance, we have
\begin{equation}
  \las \on_r (\tau) \on_0(0) \ras \sim \frac{(-1)^r}{(r^2+\tau^2)^{\alpha/2}}.
\end{equation}
for the charge correlation function in Eq.~(\ref{eq:chic}).
The rescaled charge susceptibility becomes
\begin{equation}\label{eq:chic2}
  \frac{\chi_\rho(\pi)}{L} \sim \frac{1}{L}\sum_{r} \int_0^\beta 
  \frac{d\tau}{(r^2+\tau^2)^{\alpha/2}}\,.
\end{equation}
Taking the continuum limit, and regularizing with a short-distance cutoff
$a$, we obtain
\begin{equation}
    \frac{\chi_\rho(\pi)}{L} \sim \frac{1}{L}\int_a^L d x \int_a^\beta d\tau \,
    (x^2+\tau^2)^{-\alpha/2}\,.
\end{equation}
Transforming to polar coordinates $(\rho,\phi)$, setting $\beta=L$, and
assuming $\alpha<2$ we find
\begin{equation}\label{eq:chic3}
    \frac{\chi_\rho(\pi)}{L} 
    \sim
    \frac{2\pi}{L} \int_a^L d \rho \,
    \rho^{1-\alpha}
    \sim  
    L^{1-\alpha} \left[1-\left(\frac{a}{L}\right)^{2-\alpha}\right]\,.
\end{equation}
Finally, taking the limit $a\to0$, we obtain
\begin{equation}\label{eq:chic4}
    \frac{\chi_\rho(\pi)}{L} 
    \sim  
     L^{1-\alpha}\,.
\end{equation}


%

\end{document}